\documentclass[twocolumn, trackchanges]{aastex63}
\usepackage{verbatim}
\usepackage{CJKutf8}
\usepackage{xcolor}
\submitjournal{ApJ}

\shorttitle{IR Excesses around Bright White Dwarfs from \textit{Gaia} and \textit{unWISE}}
\shortauthors{Lai et al.}

\begin{document}

\title{Infrared Excesses around Bright White Dwarfs from \textit{Gaia} and \textit{unWISE}. II}

\correspondingauthor{Samuel Lai}
\email{samuel.lai@anu.edu.au}

\author{\href{http://orcid.org/0000-0001-9372-4611}{Samuel Lai \begin{CJK}{UTF8}{gbsn}(赖民希)\end{CJK}}}
\affiliation{Gemini Observatory, NSF's NOIRLab,
670 N. A'ohoku Place
Hilo, Hawaii, 96720, USA}
\affiliation{Research School of Astronomy and Astrophysics, Australian National University, Canberra, ACT 2611, Australia}

\author{\href{http://orcid.org/0000-0003-2852-268X}{Erik Dennihy}}
\affiliation{Gemini Observatory, NSF's NOIRLab,
Casilla 603, La Serena, Chile}

\author{\href{http://orcid.org/0000-0002-8808-4282}{Siyi Xu \begin{CJK}{UTF8}{gbsn}(许\CJKfamily{bsmi}偲\CJKfamily{gbsn}艺)\end{CJK}}}
\affiliation{Gemini Observatory, NSF's NOIRLab,
670 N. A'ohoku Place
Hilo, Hawaii, 96720, USA}

\author{Atsuko Nitta}
\affiliation{Gemini Observatory, NSF's NOIRLab,
670 N. A'ohoku Place
Hilo, Hawaii, 96720, USA}

\author{Scot Kleinman}
\affiliation{Gemini Observatory, NSF's NOIRLab,
670 N. A'ohoku Place
Hilo, Hawaii, 96720, USA}

\author{S.K. Leggett}
\affiliation{Gemini Observatory, NSF's NOIRLab,
670 N. A'ohoku Place
Hilo, Hawaii, 96720, USA}

\author{Amy Bonsor}
\affiliation{Institute of Astronomy, University of Cambridge, Madingley Road, Cambridge, CB3 0HA, UK}

\author{Simon Hodgkin}
\affiliation{Institute of Astronomy, University of Cambridge, Madingley Road, Cambridge, CB3 0HA, UK}

\author{Alberto Rebassa-Mansergas}
\affiliation{Departament de F\'{i}sica, Universitat Polit\`{e}cnica de Catalunya, c/Esteve Terrades 5, 08860 Castelldefels, Spain}
\affiliation{Institute for Space Studies of Catalonia, c/Gran Capit\`{a} 2-4, Edif. Nexus 104, 08034 Barcelona, Spain}

\author{Laura K Rogers}
\affiliation{Institute of Astronomy, University of Cambridge, Madingley Road, Cambridge, CB3 0HA, UK}

\begin{abstract}
Infrared excesses around white dwarf stars indicate the presence of various astrophysical objects of interest, including companions and debris disks. In this second paper of a series, we present follow-up observations of infrared excess candidates from \textit{Gaia} and \textit{unWISE} discussed in the first paper, Paper I. We report space-based infrared photometry at 3.6 and 4.5 micron for 174 white dwarfs from the \textit{Spitzer Space Telescope} and ground-based near-infrared \textit{J}, \textit{H}, and \textit{K} photometry of 235 white dwarfs from Gemini Observatory with significant overlap between \textit{Spitzer} and Gemini observations. This data is used to confirm or rule-out the observed \textit{unWISE} infrared excess. From the \textit{unWISE}-selected candidate sample, the most promising infrared excess sample comes from both color and flux excess, which has a \textit{Spitzer} confirmation rate of 95\%. We also discuss a method to distinguish infrared excess caused by stellar or sub-stellar companions from potential dust disks. In total, we confirm the infrared excess around 62 white dwarfs, 10 of which are likely to be stellar companions. The remaining 52 bright white dwarf with infrared excess beyond two microns has the potential to double the known sample of white dwarfs with dusty exoplanetary debris disks. Follow-up high-resolution spectroscopic studies of a fraction of confirmed excess white dwarfs in this sample have discovered emission from gaseous dust disks. Additional investigations will be able to expand the parameter space from which dust disks around white dwarfs are found.

\end{abstract}


\section{Introduction}
\renewcommand*{\theHtable}{\arabic{table}} 
White dwarfs with circumstellar debris disks provide insight into the compositions of tidally disrupted exoplanetary bodies \citep{Debes2002, Jura2003, Jura_Extrasolar_Cosmochemistry}. Material from tidally disrupted planetesimals are linked to gases and solids involved in the exosolar planetary system's formation that were eventually incorporated into major and minor planetary bodies \citep{Bergin2015}. The study of white dwarfs with circumstellar exoplanetary debris disks is further informing our understanding of the formation, evolution, and disruption of minor planetary bodies \citep{Harrison_2018, Malamud_2020(2), Malamud_2020}. 

Dusty debris disks around white dwarfs are most often identified through their excess infrared radiation, though the excess can also be coming from any source cooler than the host white dwarf, including late-type stellar companions and brown dwarfs. The nominal frequency of white dwarfs with debris disks is estimated to be between 2-4\% \citep{Barber_2014, Rocchetto2015MNRAS.449..574R, wilson-unbiased-freq, Rebassa_Gaia} and the occurrence rate of detached white dwarfs with brown dwarf companions is estimated to be roughly 0.5-2.0\% with fewer than a dozen systems known to date \citep{Girven2011MNRAS.417.1210G, Steele2011MNRAS.416.2768S, Casewell_2018}. The occurrence rate of white dwarfs with M-dwarf companions is significantly greater at $28 \pm 3$\% \citep{WIREDII}. Both dusty debris disks and late type stellar companions around white dwarfs are rare and useful for studies of specific astrophysical phenomena (e.g. \citealt{Jura_Extrasolar_Cosmochemistry, Rappaport_2017, Longstaff_2019}). 

In 2018 a precision astrometric catalogue, \textit{Gaia Data Release 2}, became publicly available \citep{2018Gaia} and its data was used to construct a new catalogue of $\sim$ 260,000 high-confidence white dwarf candidates \citep{Gentile_2019WDs}. The first paper in this series, \cite{XuCandidates}, identified infrared excess candidates using a list of high-probability, bright (\textit{Gaia} $\textit{G} < 17.0$ mag) white dwarfs from \textit{Gaia DR2} and photometry from \textit{unWISE} \citep{unWISE2019}, which is a catalogue combining all of the available \textit{NEOWISE} and \textit{WISE} original epochs. Hereafter, we refer to Xu et al. 2020 as ``Paper I''. Using specific reproducible selection criteria, a sample of the best \textit{unWISE} infrared excess candidates were filtered out resulting in 188 final candidates\footnote{There was a typo in the original machine readable table of Paper I. Please use \citet{Xu_2021}.}. However, white dwarfs selected using the methodology outlined in Paper I are still affected by \textit{WISE} source confusion and contamination, which is the main limitation for \textit{WISE}-selected infrared candidates \citep{Dennihy_2020confusion}. This study presents follow-up observations of candidates from Paper I to confirm the presence of infrared excess and, in some cases, identify the likely source. 

We present infrared photometric observations of 183 targets observed with the \textit{Spitzer Space Telescope} and 235 targets observed with Gemini Observatory near-infrared imagers, with 98 observed with both. The target selection did not closely adhere to Paper I's final selection criteria, but the observed sample of white dwarfs includes 92 of the final 188 identified infrared excess candidates of Paper I, 56 of which were observed with \textit{Spitzer}. In Section \ref{sec:2-obs}, we discuss the white dwarf sample and their photometric observations. In Section \ref{sec:Spitzer_excess}, we discuss the modeling of the white dwarf photospheric flux and construct spectral energy distributions for each target. We also show the methodology for confirming the existence of infrared excess with \textit{Spitzer} and assess how the results can inform future studies of infrared excess candidates without \textit{Spitzer} data. In Section \ref{sec:disk-or-bd}, we show a method, using near-infrared photometry, which indicates when the source of the infrared excess is likely a low-mass companion rather than a dust disk. In Section \ref{sec:Discussion}, we discuss the infrared excess findings and their implications for the remaining sample of candidates from Paper I. In Section \ref{sec:conclusion}, we conclude with a summary of the results and a description of the best remaining candidates for follow-up observations.

\section{Observation and Data Reduction} \label{sec:2-obs}

The selection process of Paper I separates white dwarfs into five distinct samples (ABCDE), each according to well-defined characteristics. Each successive sample is a subset of the prior sample (see Figure 3 in Paper I). The final sample of 188 highest-confidence infrared excess candidates make up ``Sample E" in Paper I. It includes 22 known white dwarf debris disks and three known white dwarf-brown dwarf pairs. Hereafter, we refer to the final sample of 188 infrared excess candidates from Paper I as ``Sample E". All observed white dwarfs outside of Sample E are referred to as ``Samples A-E", meaning Sample A through to Sample D. Target selection for this study was largely independent of Paper I's selection criteria. For \textit{Spitzer} observations, we selected targets from a preliminary infrared excess candidate list compiled before the methodology of Paper I was fully established. For near-infrared imaging, we selected targets based on lack of publicly available \textit{J}, \textit{H}, and \textit{K} photometry. As a result, the white dwarfs in this study were sourced from every Sample as defined in Paper I. Table \ref{tab:observed_sample} shows the breakdown of observed white dwarfs by their original Sample in Paper I. 

\begingroup
\squeezetable
\begin{table}
\caption {\label{tab:observed_sample} Number of targets observed by \textit{Spitzer's IRAC} and Gemini near-infrared instruments categorised into Samples A-E and Sample E as described in Paper I. The second column indicates the total number of white dwarfs in each Sample.} 
\begin{ruledtabular}
\begin{tabular}{cccc}
 Sample & Paper I & \textit{Spitzer} & Gemini (\textit{NIRI}/\textit{F2}) \\    \hline
 Sample A-E & 5814 & 118 &  165 (61/108) \\
 Sample E & 188 & 56 & 70 (55/18)
\end{tabular}
\end{ruledtabular}
\end{table}
\endgroup

\subsection{IRAC Imaging and Photometry}
We observed 183 targets with the \textit{InfraRed Array Camera} (\textit{IRAC}) in both warm channels with central wavelengths located at 3.6 (\textit{Ch1}) and 4.5 microns (\textit{Ch2}) respectively \citep{Spitzer_orig} under the programme number 14220. Each exposure was 30 seconds and 11 medium-sized dithers were used for each wavelength. Both point-response function (PRF) and aperture photometry were performed for every observation using the \textit{MOsaicker and Point source EXtractor} (\textit{MOPEX}) package. The measurement flux uncertainty was added in quadrature with an additional calibration uncertainty of 5\% \citep{Farihi_2008}. We report the PRF magnitudes unless the target did not appear properly subtracted in the residual image or the centroid position of the target between the \textit{Ch1} and \textit{Ch2} frames was significantly discrepant. In those cases, we reported magnitudes measured from aperture photometry. Observed targets have a typical measurement signal-to-noise ratio (SNR) of 17.5 in \textit{Ch1} and 18.2 in \textit{Ch2} after applying the systematic calibration uncertainty.

As they were selected for having \textit{unWISE} excess, many of our targets are found in crowded fields with the risk of contamination from nearby sources, even in the higher spatial resolution \textit{Spitzer} images. The PRF-fitted photometry can mitigate this risk, but is not immune to cases of overlapping sources or nearby extended sources. The reliability of the \textit{Spitzer} photometry was determined by examination of the PRF residual images, which we searched for evidence of over- or under-subtraction of the target source. We have flagged 9 of 183 targets for which the PRF was not cleanly subtracted and this \textit{Spitzer} flag is indicated by the letter ``s'' in Table \ref{tab:no-excess_all}. Photometry of targets with the \textit{Spitzer} flag is considered unreliable and any indication of excess is not likely to be real. We exclude flagged targets from excess statistics but report them in Table \ref{tab:no-excess_all} for completeness. The remaining 174 targets with reliable \textit{Spitzer} photometry are examined for infrared excess in Section \ref{sec:assess-excess}.

\subsection{\textit{NIRI} and \textit{FLAMINGOS-2} Imaging and Photometry}
We obtained new near-infrared photometry for a total of 116 white dwarfs using the \textit{Near InfraRed Imager} (\textit{NIRI}; \citealt{NIRI}) at Gemini North and 126 targets from \textit{Flamingos-2} (\textit{F2}; \citealt{F2}) at Gemini South. In total, 235 white dwarfs were observed, accounting for some overlap between \textit{NIRI} and \textit{F2}. Observations were conducted under a variety of weather conditions. More information on the Gemini observations and data processing can be found in the \nameref{Appendix} section of the Appendix. A small sample of the infrared photometry of targets observed by both Gemini and \textit{Spitzer} is available in Table \ref{tab:photometry} with the full table containing all of the new photometry available in digital form\footnote{Temporarily available at \url{https://www.mso.anu.edu.au/~samlai/Table_2_MR.csv}}.

\textit{Gemini-N/NIRI}: We conducted observations of 116 white dwarfs using \textit{NIRI} at Gemini North as part of programs GN-2018B-FT-208, GN-2018B-Q-406, GN-2019A-Q-303, GN-2019A-Q-403, GN-2019B-FT-111, GN-2019B-FT-216, GN-2019B-Q-237,  GN-2019B-Q-408, and GN-2020A-Q-405. Each target was observed in the Mauna Kea Observatory (\textit{MKO}) standard \textit{J}, \textit{H}, and \textit{K} filters within a 120"x 120" field of view. Exposures were 10 seconds and a random dither pattern was employed for a total of approximately 20 exposures in each filter. Data reduction and frame-stacking were handled by version 2.1.0 of Gemini's publicly available Data Reduction for Astronomy from Gemini North and South (\textit{DRAGONS}) software \citep{DRAGONS}. Aperture photometry was performed with \textit{astropy}'s \textit{photutils} \citep{astropy} using its point-spread function (PSF) fitting capabilities. The reference stars were modeled independently to determine an appropriate aperture radius for each frame based on three times the median of the full-width half-maximum (FWHM) determined by the PSF fitting. Bright reference stars in the \textit{Two Micron All-Sky Survey} (\textit{2MASS}; \citealt{2MASS}), where the \textit{2MASS $K_{\rm{s}}$} $\leq$ 15.5 mag, located within the field of view, were used to calibrate the zero-points for all \textit{J}, \textit{H}, and \textit{K} bandpasses. For \textit{J} and \textit{H} images, we used the same bright reference stars from \textit{K}-band unless the total number of reference stars is less than five. In which case, we used all of the \textit{2MASS} stars contained within the field of view.  The \textit{2MASS} photometry of reference stars was converted into the \textit{MKO} photometric system in order to calibrate a static flux zeropoint for each image \citep{2mass-mko-conversion}. The typical SNR for our targets were 170, 170, and 140 for \textit{J}, \textit{H}, and \textit{K} bands respectively.

\textit{Gemini-S/F2}:   For targets in the southern sky, we conducted observations of 126 white dwarfs at Gemini South using \textit{F2} as part of programs GS-2018B-FT-204, GS-2018B-Q-404, GS-2019A-Q-301, GS-2019A-Q-404, GS-2019B-Q-237, GS-2019B-Q-408, and GS-2020A-Q-409. Seven of the observed targets were also observed with \textit{NIRI}. Each target was observed in the \textit{MKO J} and \textit{H} bandpasses, as well as the \textit{$K_s$} bandpass, with a 6' circular field of view. As with \textit{NIRI}, we used a random dither pattern around the target, but the exposure times were often between 10-30 seconds for \textit{J} band, 6-10 seconds for \textit{H} band, and 10-20 seconds for \textit{$K_s$} band. Upwards of 20 exposures per target in each filter were taken. We used \textit{DRAGONS} for data reduction and performed aperture photometry in the same way as with targets observed by \textit{NIRI} using bright \textit{2MASS} reference stars within the field of view. Since the \textit{F2} \textit{$K_s$} filter is similar to that of \textit{2MASS} \citep{sandy-nir-dwarfs}, no transformation was made for the \textit{$\rm{K_s}$} magnitude to convert it into the \textit{MKO} system preceding the zeropoint calibration. Typical SNR across all \textit{F2} targets and filters were 180, 140, and 110 for \textit{J}, \textit{H}, and \textit{$K_s$} bands respectively. 

Some of the white dwarfs observed with Gemini have existing near-infrared photometry. The Gemini photometry was compared against the \textit{UKIRT Infrared Deep Sky Survey} (\textit{UKIDSS}) for \textit{NIRI} and \textit{VISTA Hemisphere Survey} (\textit{VHS}) for \textit{F2}. The comparisons showed systematic linear offsets in the \textit{J} band for \textit{NIRI}, and in the \textit{J} and \textit{$K_s$} bands for \textit{F2}. We measured and corrected for this offset. Additional details, including the magnitude of the offsets, are shown in the Appendix Section \ref{appendix:filter_transformations}.

The Gemini aperture photometry can be unreliable for reasons including blending with background objects or poor zeropoint calibration. We found the uncertainty of our photometric results to increase up to 30\% when the number of quality reference stars within the field of view is small, thus we have flagged all photometry where four or fewer reference stars were used. In cases where the Gemini photometry is flagged, we consider public photometry from \textit{UKIDSS}, \textit{UKIRT Hemisphere Survey} (\textit{UHS}), and \textit{VISTA} instead for calculations involving near-infrared excess if they exist. The \textit{VISTA} photometry is transformed into the \textit{MKO} system. If no public near-infrared photometry is available, we use the flagged Gemini photometry as a last resort. In total, 81 targets were flagged out of 235 targets observed with Gemini Observatory. Flagged targets should be treated with additional caution. 

\begin{figure*}[ht!]
\plotone{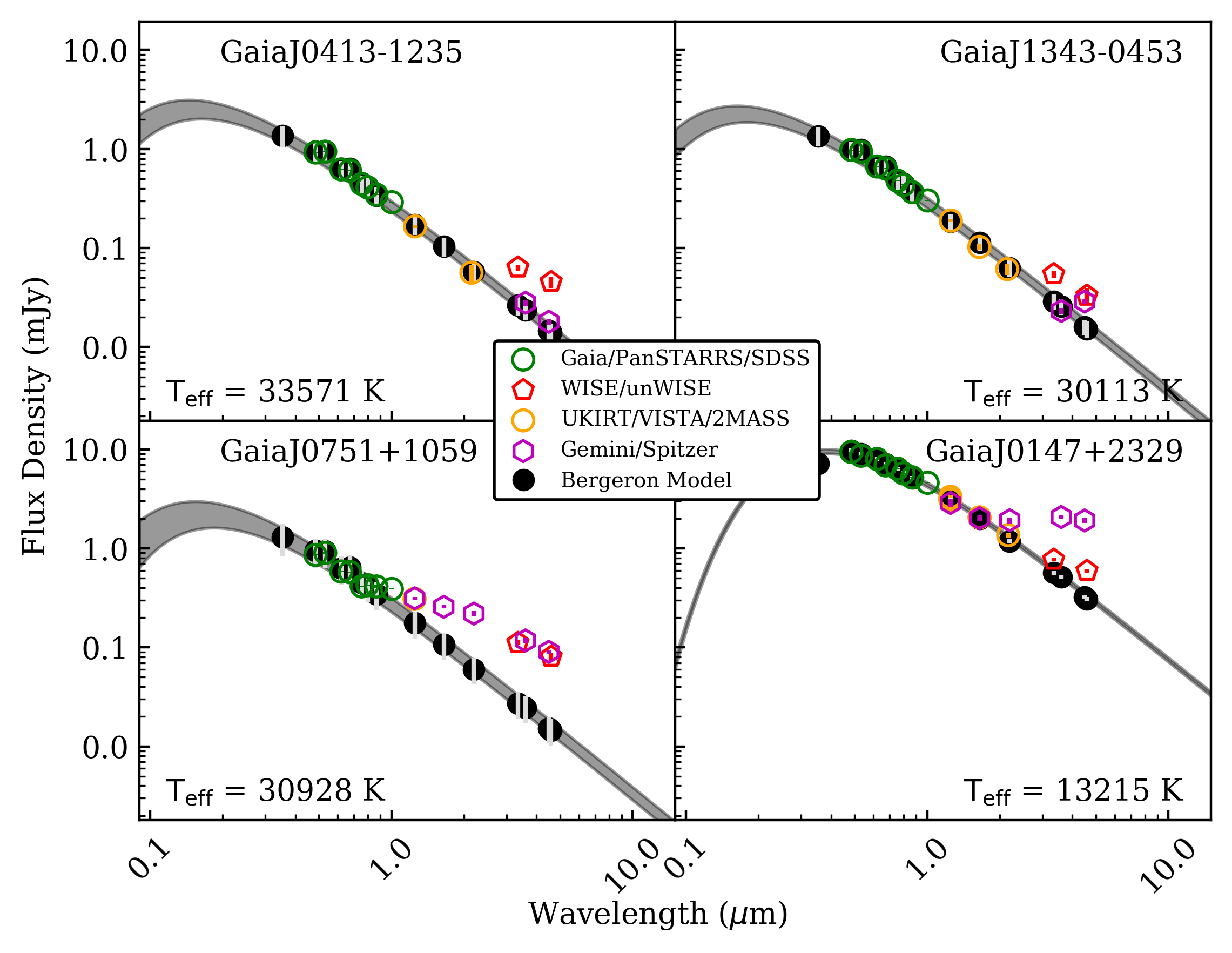}
\caption{Selected spectral energy distributions of our targets. For visual purposes, the photometry is fitted with two photometric models: the black points mark the photometric Bergeron model \citep{Bergeron_model} based on atmospheric parameters determined in \cite{Gentile_2019WDs}, and the shaded region traces a blackbody model and its one-sigma uncertainty. The SED on the top left panel (GaiaJ0413-1235) shows a case of an infrared excess candidate from Paper I that is ruled out by the higher quality \textit{Spitzer} photometry. These cases are the result of source confusion in \textit{unWISE} that are resolved by \textit{Spitzer}. The top-right panel (GaiaJ1343-0453) shows one target with color-only excess. The bottom-left panel (GaiaJ0751+1059) shows an example of an infrared excess characteristic of a white dwarf with a low-mass companion and the bottom-right panel (GaiaJ0147+2329) shows infrared excess from a known circumstellar dust disk. GaiaJ0147+2329 is also a special case of a disk with high infrared variability, with over one magnitude of difference between the \textit{Spitzer} and \textit{unWISE} photometry, due to a possible tidal disruption in progress \citep{irvariable}. All of the SEDs are available online.} \label{fig:SED_grid}
\end{figure*}

\begingroup
\squeezetable
\begin{table*}[t]
\caption {\label{tab:photometry} Photometry of white dwarfs observed by both \textit{Spitzer} and one of Gemini's near-infrared instruments, \textit{NIRI} or \textit{F2}. Gemini near-infrared \textit{J}, \textit{H}, \textit{K} magnitudes are reported in the \textit{MKO} system. Color transformations have been applied to \textit{F2 $K_s$}. Photometry flag, ``g'', indicates when the Gemini photometry is based on a low number of reference stars, which is also correlated with higher uncertainty. The total number of targets observed is 320, with near-infrared photometry reported for 180 targets for the first time and new \textit{Spitzer} photometry for 174 targets.}  
\begin{ruledtabular}
\begin{tabular}{lllccccccc}
 Name & \textit{Gaia} RA & \textit{Gaia} Dec & \textit{J} & \textit{H} & \textit{K} & Instrument & \textit{Spitzer Ch1} & \textit{Spitzer Ch2}  & Flags  \\ 
      & (deg) & (deg) & (mag)  & (mag) & (mag) & & (mag) & (mag)                                                              \\    \hline
GaiaJ0428+3644 & 67.078300 & 36.739478 & 17.21 $\pm$ 0.06 & 17.27 $\pm$ 0.03 & 17.28 $\pm$ 0.03 & \textit{NIRI} & 17.25 $\pm$ 0.07 & 16.91 $\pm$ 0.07 &  \\ 
GaiaJ0609+3913 & 92.250011 & 39.222588 & 17.58 $\pm$ 0.05 & 17.73 $\pm$ 0.03 & 17.77 $\pm$ 0.03 & \textit{NIRI} & 17.62 $\pm$ 0.08 & 17.68 $\pm$ 0.08 &  \\ 
GaiaJ0834+5336 & 128.588430 & 53.604311 & 17.44 $\pm$ 0.05 & 17.68 $\pm$ 0.04 & 17.74 $\pm$ 0.03 & \textit{NIRI} & 17.67 $\pm$ 0.09 & 17.74 $\pm$ 0.08 &  \\ 
GaiaJ0902+3120 & 135.677408 & 31.345378 & 15.14 $\pm$ 0.11 & 15.09 $\pm$ 0.14 & 15.19 $\pm$ 0.15 & \textit{NIRI} & 15.07 $\pm$ 0.06 & 15.07 $\pm$ 0.06 & g \\ 
GaiaJ1903+6035 & 285.833014 & 60.598328 & 15.20 $\pm$ 0.07 & 15.25 $\pm$ 0.05 & 15.04 $\pm$ 0.06 & \textit{NIRI} & 14.05 $\pm$ 0.06 & 13.51 $\pm$ 0.06 &  \\ 
GaiaJ2233+8408 & 338.321327 & 84.137396 & 16.37 $\pm$ 0.06 & 16.42 $\pm$ 0.04 & 16.46 $\pm$ 0.04 & \textit{NIRI} & 16.43 $\pm$ 0.06 & 16.51 $\pm$ 0.06 &  \\ \hline
GaiaJ0107+2518 & 16.859511 & 25.309778 & 17.07 $\pm$ 0.05 & 17.12 $\pm$ 0.03 & 17.26 $\pm$ 0.07 & \textit{F2} & 17.23 $\pm$ 0.07 & 17.09 $\pm$ 0.07 &  \\ 
GaiaJ0347+1624 & 56.902909 & 16.402432 & 16.92 $\pm$ 0.05 & 16.88 $\pm$ 0.04 & 16.60 $\pm$ 0.07 & \textit{F2} & 16.40 $\pm$ 0.06 & 15.90 $\pm$ 0.06 &  \\ 
GaiaJ0421+1529 & 65.453585 & 15.487452 & 17.04 $\pm$ 0.06 & 17.11 $\pm$ 0.03 & 17.10 $\pm$ 0.08 & \textit{F2} & 17.04 $\pm$ 0.07 & 16.98 $\pm$ 0.07 &  \\ 
GaiaJ0950+1837 & 147.529107 & 18.625792 & 16.82 $\pm$ 0.05 & 16.83 $\pm$ 0.03 & 16.95 $\pm$ 0.09 & \textit{F2} & 16.81 $\pm$ 0.06 & 16.90 $\pm$ 0.07 &  \\ 
GaiaJ1155+2649 & 178.775677 & 26.823271 & 17.17 $\pm$ 0.07 & 17.19 $\pm$ 0.07 & 17.33 $\pm$ 0.08 & \textit{F2} & 16.85 $\pm$ 0.06 & 16.72 $\pm$ 0.06 &  \\ 
GaiaJ1449-3029 & 222.388494 & -30.488730 & 16.41 $\pm$ 0.05 & 16.29 $\pm$ 0.03 & 16.13 $\pm$ 0.08 & \textit{F2} & 16.10 $\pm$ 0.06 & 16.07 $\pm$ 0.06 &  \\ 
GaiaJ1612+1419 & 243.026837 & 14.318464 & 16.51 $\pm$ 0.07 & 16.49 $\pm$ 0.02 & 16.63 $\pm$ 0.12 & \textit{F2} & 16.50 $\pm$ 0.06 & 16.51 $\pm$ 0.06 & g \\ 

\end{tabular}
\end{ruledtabular}
\begin{tabbing}
$^{\rm{g}}$Gemini photometry flag. \\
Note: This table is available in its entirety in machine-readable format .
\end{tabbing}
\end{table*}
\endgroup

\section{Confirming the Infrared Excess with \textit{Spitzer}} \label{sec:Spitzer_excess}

In the previous section, we presented the new infrared photometry of 235 white dwarfs using Gemini Observatory and 174 white dwarfs using the \textit{Spitzer Space Telescope}. With its higher spatial resolution and sensitivity, the \textit{Spitzer} photometry is ideally suited to confirm or rule-out the presence of infrared excess in the candidates of Paper I. In this section, we construct spectral energy distributions (SEDs) of all targets observed by \textit{Spitzer} and compare them with stellar models. We also establish quantitative metrics of color and flux excess. These metrics are applied to the new \textit{Spitzer} and Gemini observations to confirm or rule-out the observed infrared excess.

\subsection{Stellar Model Comparisons}
SEDs were constructed using photometry from the \textit{Panoramic Survey Telescope and Rapid Response System} (\textit{PanSTARRS}; \citealt{panstarrs}) DR1, \textit{Sloan Digital Sky Survey} (\textit{SDSS}; \citealt{SDSS}) DR12, \textit{VHS} \citep{VISTA_Archive} DR6, \textit{UKIDSS} \citep{UKIDSS, WFCAM_Archive} DR11, \textit{UHS} \citep{UHS_DR1} DR1, \textit{2MASS} \citep{2MASS}, \textit{ALLWISE} \citep{ALLWISE}, and \textit{unWISE} \citep{unWISE2019}. For our white dwarf photometric and spectroscopic models, we assume the effective temperature, $T_{\rm{eff}}$, and surface gravity of the DA model fits reported in Gentile Fusillo \citep{Gentile_2019WDs}. For each target, synthetic photometry of a DA white dwarf \citep{Bergeron_model}\footnote{\url{https://www.astro.umontreal.ca/~bergeron/CoolingModels/}} was scaled to fit the \textit{PanSTARRS}, \textit{SDSS}, or \textit{Gaia} optical photometry using chi-square minimisation methods. Hereafter, we refer to the DA white dwarf synthetic photometry as the ``Bergeron model''. For visual purposes, a blackbody model was adjusted to fit the \textit{J} flux density of the photometric Bergeron model in the SED figures. The model flux and its corresponding uncertainty described in the following sections refer to the photometric Bergeron model. 

All photometric magnitudes were converted into flux densities using associated zero-points for each bandpass. Gemini photometry was converted into flux density using photometric zeropoints of the \textit{MKO} system \citep{2mass-mko-conversion}. Filter transformation is described in detail in the Appendix Section \ref{appendix:filter_transformations}.

Sources of statistical error in the model include the uncertainty in the temperature, parallax, and surface gravity. The model error was computed by fitting a Poissonian probability distribution to a set of apparent magnitudes generated from a Monte-Carlo simulation. For each magnitude, the temperature was randomly sampled from a Gaussian distribution of its mean and standard deviation. A resulting magnitude was obtained from each set of white dwarf parameters by referring to the pure-hydrogen model white dwarf atmosphere grid from the Bergeron model. We applied an uncertainty floor of 5\% in the model flux to represent the systematic uncertainty of fitting every target with a DA white dwarf model while lacking information on each individual target's spectral type. If the white dwarf parameters are off or it is a different spectral type than the assumed DA, it can lead to an erroneous stellar temperature affecting the predicted infrared flux in the bands of interest. As a secondary check, we have also performed the infrared stellar model flux calculations for all targets using DB parameters from Gentile Fusillo et. al. and DB models from Bergeron, finding a median flux difference of 4\% although other white dwarf models may result in a larger discrepancy. A sample of representative SEDs are shown in Figure~\ref{fig:SED_grid} and the rest are submitted as digital content\footnote{Temporarily available at \url{https://www.mso.anu.edu.au/~samlai/SEDs.zip}}.

\subsection{Assessing Infrared Excess} \label{sec:assess-excess}
With the new infrared data presented in Section \ref{sec:2-obs} and the Bergeron stellar model fluxes \citep{Bergeron_model}, we are able to re-assess the infrared excess for each target. Following the methods outlined in Paper I, we use both the flux excess metric and mid-infrared color excess metric to search for infrared excesses. We define the flux excess metric, $\chi_{\rm{i}}$, as
\begin{equation}
    \chi_{\rm{i}} = \frac{F_{\rm{obs,i}} - F_{\rm{mod,i}}}{\sqrt{\sigma^2_{\rm{obs,i}} + \sigma^2_{\rm{mod,i}}}} \,,
    \label{eq:flux-excess}
\end{equation}
where the indices, $i$, indicate any single band-pass. $F_{\rm{obs,i}}$ and $F_{\rm{mod,i}}$ are the observed flux and model flux respectively, while $\sigma$ indicates the flux error. An alternative version of the flux excess metric used in Paper I measures magnitude excess with the observed and model magnitudes as well as their errors. In our sample, there was no significant difference between the flux and magnitude excess. Many of the observed targets, including all white dwarfs in Sample E, were identified in Paper I with both \textit{unWISE} magnitude excess metrics exceeding the thresholds $\chi_{\rm{W1}} > 5$ and $\chi_{\rm{W2}} > 5$. We calculated \textit{Spitzer} $\chi_{\rm{ch1}}$ and $\chi_{\rm{ch2}}$ for 56 infrared excess candidates in Sample E and 118 white dwarfs in Sample A-E. We also measured Gemini $\chi_{\rm{J}}$, $\chi_{\rm{H}}$, and $\chi_{\rm{K}}$ values for 70 infrared excess candidates in Sample E and 165 white dwarfs in Sample A-E with significant overlap between targets observed by \textit{Spitzer} and Gemini. We show the excess \textit{Spitzer} $\chi_{\rm{ch1}}$ and $\chi_{\rm{ch2}}$ values for all of observed candidates in Tables \ref{tab:excess_all} and \ref{tab:no-excess_all}.

\begin{figure*}[ht!]
\plotone{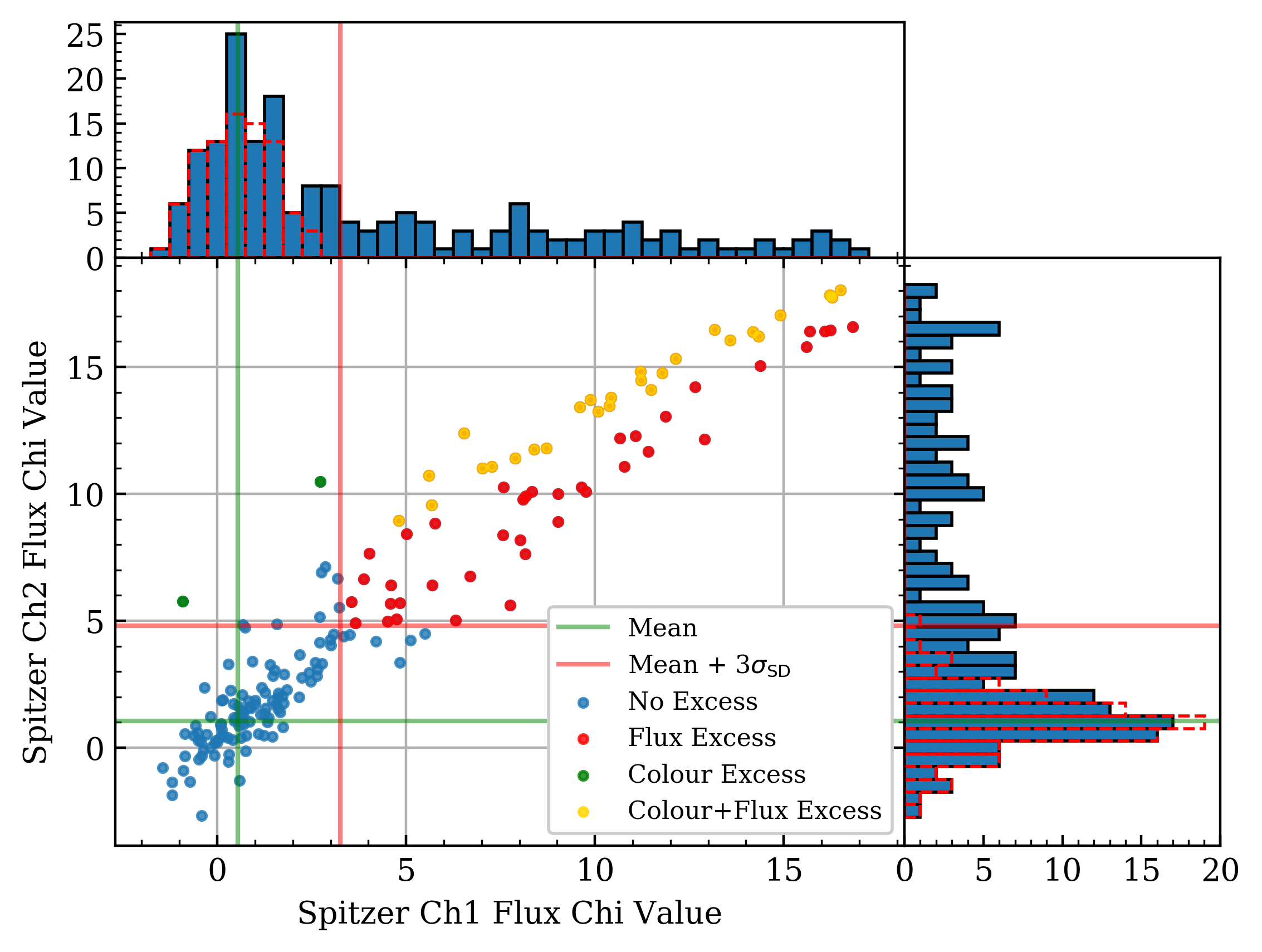}
\caption{Distribution of flux excess $\chi_{\rm{i}}$ values for \textit{IRAC Ch1} and \textit{Ch2} for the \textit{Spitzer}-observed targets with GaiaJ2012-5957 removed due to poor model fit. The 3-sigma threshold using the statistical distribution from our observed sample is shown as a solid red line and the mean is shown in green. The locus of points around the mean is likely dominated by false positive \textit{unWISE} excess candidates that were the result of contamination or source confusion. These cases were resolved by \textit{Spitzer}, but some residual contamination in our \textit{Spitzer} photometry leads to the small offset of the locus from zero. The 3-sigma threshold takes the offset from zero into account for the determination of statistically significant flux excess. Targets with \textit{Spitzer} flux excess exceeding the designated threshold are plotted in red and the mirrored distribution used to calculate the standard deviation is shown by the red-dashed histogram. Plotted in purple are \textit{Spitzer}-observed white dwarfs with both color and flux excess and color-only excess targets are plotted in navy blue. \label{fig:Spitzer_chi}}
\end{figure*}

In Figure~\ref{fig:Spitzer_chi}, we show the distribution of \textit{Spitzer} \textit{Ch1} and \textit{Ch2} chi-values for all of the observed targets. There is a locus of points clustered around the chi values of $\langle \chi\rangle_{\rm{ch1}} = 0.54 \pm 0.91$ and $\langle \chi \rangle_{\rm{ch2}} = 1.05 \pm 1.25$, as indicated by the solid green lines. This offset from zero is not likely due to real excess, but rather originating from systematics of the photometry or source confusion. As discussed in Paper I, many of the \textit{unWISE} candidates are likely false positives, with the \textit{unWISE} excess being the result of nearby unresolved objects. Though the \textit{Spitzer} data has much higher quality, we still expect some contamination from nearby sources in our PRF fluxes, which could be contributing to the offsets. To identify targets with true flux excess, we assume the negative dispersion around the locus is due to statistical fluctuation. A Gaussian profile is fitted to a synthetic distribution created by mirroring the negative dispersion across the most populated bin. The mirrored distribution is shown by the red dashed histogram outline in Figure \ref{fig:Spitzer_chi}. We consider infrared excess to be statistically significant as observed by \textit{Spitzer} if $\chi_{\rm{i}} > \left(3\sigma + \rm{offset}\right)$ in both \textit{IRAC} \textit{Ch1} and \textit{Ch2}. Therefore, the \textit{Spitzer} flux excess thresholds are $\chi_{\rm{ch1}} > 3.27$ and $\chi_{\rm{ch2}} > 4.80$ and both conditions must be satisfied for a target to have confirmed flux excess. We found 43 targets with flux excess in Sample E and 17 targets with flux excess in Samples A-E. 

The mid-infrared single-color excess, $\Sigma_{\rm{ch1-ch2}}$, hereafter shortened to ``color excess'', is defined for the two warm bandpasses \textit{Spitzer Ch1} and \textit{Ch2} as
\begin{equation}
    \Sigma_{\rm{ch1-ch2}} = \frac{m_{\rm{obs,ch1}} - m_{\rm{obs,ch2}} - m_{\rm{mod,ch1-ch2}}}{\sqrt{\sigma^2_{\rm{obs,ch1}} + \sigma^2_{\rm{obs,ch2}} + \sigma^2_{\rm{mod,ch1-ch2}}}} \,,
\end{equation}
where all quantities and uncertainties are in units of magnitude. The numerator measures the difference between the observed and model color, which is normalised against their uncertainties added in quadrature. In Paper I, we considered a target to have color excess if $\Sigma_{\rm{W1-W2}} > 3$. For our \textit{Spitzer}-observed sample, we found the mean color-excess metric $\langle{\Sigma}\rangle_{\rm{ch1-ch2}} = 0.31 \pm 0.59$. We minimised the effect of the infrared-excess bias in our \textit{Spitzer}-observed sample by removing targets where flux excess is observed in either \textit{Ch1} or \textit{Ch2} from the mean color-excess metric estimate. Using the same $\Sigma > 3$ threshold consistent with Paper I, we find 23 candidates in Sample E and five targets in Samples A-E with color excess, all but two of which also host a flux excess. As seen in Figure \ref{fig:Spitzer_chi}, roughly 40\% of observed white dwarfs in our sample exhibiting flux excess also have color excess. We show the full table of all white dwarfs with \textit{Spitzer} infrared excess and a table of all \textit{Spitzer}-observed white dwarfs without infrared excess in the Appendix, Tables \ref{tab:excess_all} and \ref{tab:no-excess_all}. 

In this paper, white dwarfs are considered to have an IR excess when either flux excess or color excess metric exceed the threshold. In this study, we have identified a total of 62 systems with \textit{Spitzer}-confirmed infrared excess across all Samples ABCDE, 44 of which are from the \textit{unWISE}-selected candidates in Sample E originally from Paper I.

\section{Separating Companions from Dust Disks with Near-Infrared Photometry} \label{sec:disk-or-bd}

In this section, we discuss one application of the near-infrared photometry in distinguishing between two potential infrared excess sources: stellar companions and circumstellar dust disks/brown dwarfs. The Gemini photometry provides the critical near-infrared flux measurement where the infrared excess from a companion or a dust disk can potentially be disentangled.

The source of the white dwarf infrared excess identified in this study can originate from companions or dust disks of varying temperatures. In the warm \textit{Spitzer} and \textit{unWISE} bandpasses, these two cases are difficult to distinguish (Figure 1 in Paper I). However, known dust disks typically do not show significant infrared excess at wavelengths shorter than two micron \citep{Farihi2016}. This is consistent with the expectation that the inner edges of the dust disk are terminated at temperatures between 1600K to 2000K \citep{Rafikov_2012}. In contrast, low-mass stellar companions span a much larger range of temperatures. As such, high-quality near-infrared photometry can be used to determine when the infrared excess source is not likely due to the presence of a dust disk \citep{Barber_2014}. The lack of excess shorter than two micron is not sufficient to confirm the presence of dusty debris or rule out companions altogether, but a confirmation of excess in the near-infrared \textit{J} bandpass can be used to rule out dust disks as the source of the excess.

\begin{figure*}[ht!]
\plotone{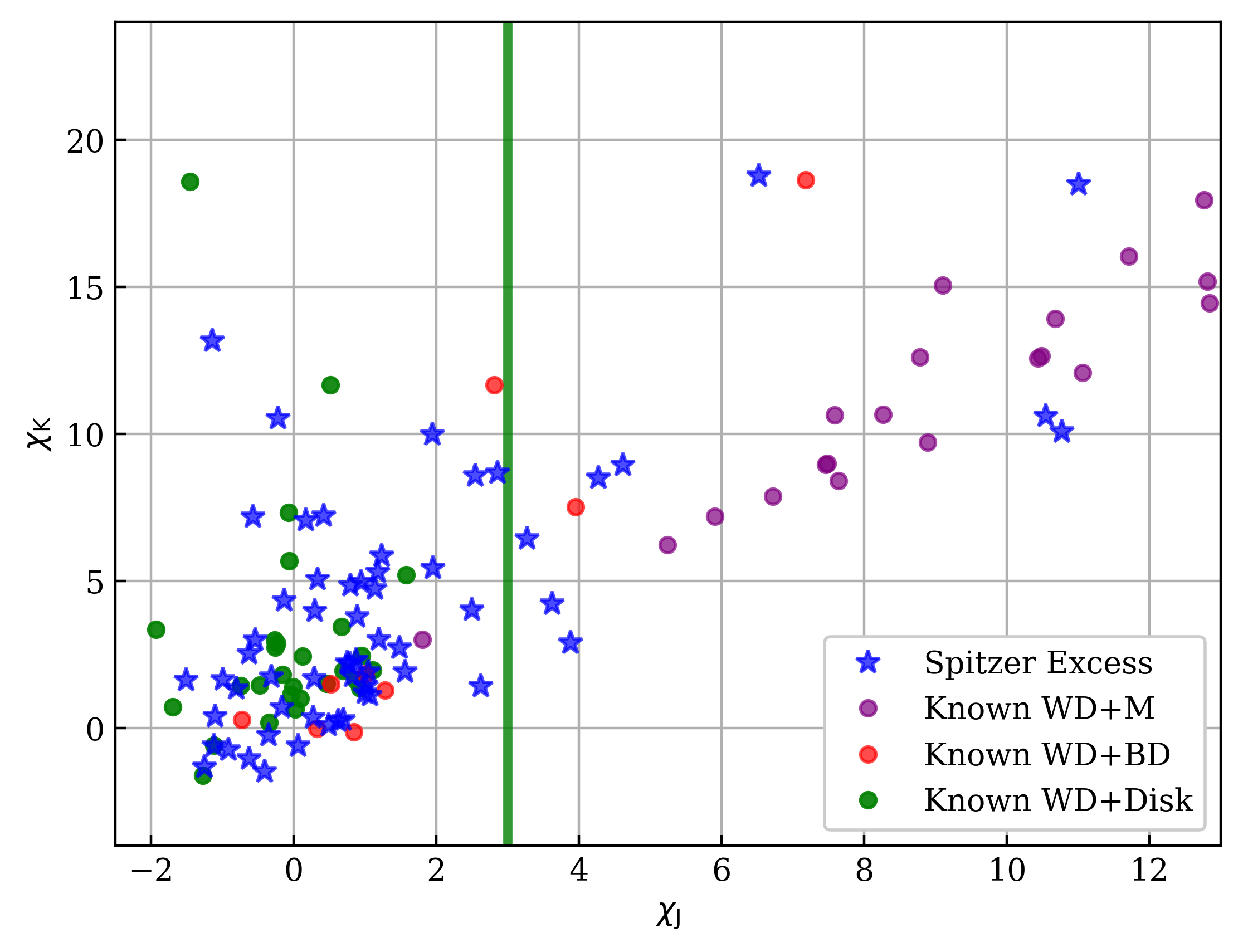}
\caption{Observed candidates' $\chi_{\rm{J}}$ and $\chi_{\rm{K}}$ metrics compared against a sample of known white dwarfs with brown dwarf companions (WD+BD), a sample of known white dwarfs with M-dwarf companions (WD+M), and a sample of known white dwarfs with circumstellar dust disks (WD+Disk). White dwarfs in binary systems with low-mass companions are sometimes differentiated by higher $\chi_{\rm{J}}$, while no known white dwarfs with circumstellar dust disks exhibit high $\chi_{\rm{J}}$. The green solid vertical line denotes a $\chi_{\rm{J}} = 3$ threshold. Any \textit{Spitzer} or \textit{unWISE} excess candidates that lie beyond the $\chi_{\rm{J}}$ line are likely to host excess from a stellar or sub-stellar companion rather than a dust disk. \label{fig:chiJ-chiK}}
\end{figure*}

The flux excess metric in Eq.~\ref{eq:flux-excess} is used to quantify the magnitude of the excess in the near-infrared bands. In Figure \ref{fig:chiJ-chiK}, our candidates are compared against the near-infrared photometry of a known sample of dusty white dwarfs (WD+Disk), a sample of known white dwarf with brown dwarf (WD+BD) binaries \citep{BD1, BD2, BD3, Casewell_2018}, and a sample of known white dwarfs with M-dwarf (WD+M) companions \citep{2016Rebassa_WDM}. The known sample of dusty white dwarfs is found in Table 1 of Paper I. The frequency of WD+M systems is low for this sample as the selection criteria discussed in Paper I avoided propagating most WD+M systems into the final sample of candidates.

While some dusty white dwarfs exhibit high $\chi_{\rm{K}}$, most of the M-dwarf companion systems exhibit significant $\chi_{\rm{J}}$ as well, which is uncommon for dust disks. The surface gravity of many binary white dwarf systems with low $\chi_{\rm{J}}$ is poorly constrained, which affects the model uncertainty and significantly lowers the $\chi_{\rm{J}}$ value. This is not the case for the known white dwarfs with dust disks. The mean $\chi_{\rm{J}}$ for the population of known white dwarfs with circumstellar dust disks is $\langle \chi\rangle_{\rm{J}} = -0.03 \pm 0.89$. We use this metric to identify \textit{Spitzer}-confirmed infrared excess candidates with $\chi_{\rm{J}} \geq 3.0$ as likely to be hosting a companion rather than a dust disk. This threshold is indicated in Figure \ref{fig:chiJ-chiK} by the solid green vertical line. The $\chi_{\rm{J}}$ value of known white dwarfs with dust disks is consistent with zero, such that the infrared excess of targets beyond the $\chi_{\rm{J}} = 3.0$ threshold is unlikely to be the result of a dusty debris disk. We display this property with a $\chi_{\rm{J}}$ flag when reporting the targets with \textit{Spitzer} infrared excess in Table \ref{tab:excess_all} of the Appendix. Out of the 62 total targets with Spitzer excess, all but two have some existing high-quality near-infrared photometry from public surveys or new Gemini photometry. A total of 10 targets are flagged with a high $\chi_{\rm{J}} > 3.0$ value, indicating likely stellar or substellar companion excess. Four of these, GaiaJ0007+1951, GaiaJ0751+1059, GaiaJ1131-1438, and GaiaJ1731-1002, also have significant \textit{PanSTARRS y}-band excess over 3$\sigma$ which could be indicative of a low-mass stellar companion over a brown dwarf. Additional caution should be applied for one target, GaiaJ0507+4541, as its Gemini photometry is flagged and no public near-infrared photometry is used in its place. Near-infrared spectroscopy and time series photometry would be useful to confirm the origin of the infrared excess.

\section{Discussion} \label{sec:Discussion}
In this section, we discuss the targets with \textit{Spitzer}-confirmed infrared excesses. We present a summary of targets with \textit{Spitzer} excess in Table \ref{tab:confirmed_sample} with the full table of infrared excess targets in the Appendix, Table \ref{tab:excess_all}. We also report all targets without evidence of \textit{Spitzer} excess in the Appendix, Table \ref{tab:no-excess_all}. We evaluate the implications of the results presented here on the remainder of infrared excess candidates in Paper I. In addition, we revisit our remaining candidates from Paper I that do not have \textit{Spitzer} photometry and present nine targets with the highest likelihood of exhibiting real excess for observational follow-up.

\subsection{\textit{Spitzer}-Confirmed Infrared Excesses (Sample E)} \label{sec:confirmed_excess}
We have observed 56 \textit{unWISE}-selected infrared excess candidates from the final sample of 188 candidates which makes up Sample E in Paper I.  With the excess metrics outlined in previous sections and the new \textit{Spitzer} photometry, we find 44 targets from Sample E with either flux or color excess, of which 22 satisfy both excess thresholds. Previous studies found that over 90\% of the known sample of white dwarfs with circumstellar dust disks exhibit both statistically significant flux and color excess \citep{wilson-unbiased-freq}. 

In Sample E, we find 21 stars with only flux excess and one star with color-only excess. We show the number of stars with infrared excess in Table \ref{tab:confirmed_sample}. The confirmation rate for \textit{Spitzer} excess in the 56 observed Sample E targets is 79\%. We also find seven targets with significant \textit{J}-band excess, indicating that the white dwarf is likely to have a low-mass companion.

Out of the observed Sample E white dwarfs, we found three white dwarfs with flux excess in only \textit{Spitzer Ch2} bandpass, including the one star with only color excess. This scenario could indicate the presence of cooler dust, similar to the case of HS 2132+0941 \citep{cool_excess_2014}. For the 12 \textit{Spitzer}-observed Sample E candidates without a \textit{Spitzer} flag or confirmation of excess, two of the white dwarfs exhibited only \textit{Ch2} excess and did not satisfy either of the flux or color excess thresholds. All 12 candidates exhibited \textit{unWISE} excess which is ruled out by the higher quality \textit{Spitzer} photometry. The \textit{Spitzer} residuals show one or more distinct nearby sources which were captured and blended within the large beam size of \textit{unWISE} and the \textit{unWISE}-excess was the result of source confusion. These limitations were resolved using the new \textit{Spitzer} photometry. We show an example of such a case in the top-left panel of Figure ~\ref{fig:SED_grid}.

\begingroup
\squeezetable
\begin{table}
\caption {\label{tab:confirmed_sample} Number of targets confirmed by \textit{Spitzer} with either the color or flux excess metric categorised into Samples A-E and Sample E as described in Paper I. The number of confirmed white dwarfs are further subdivided into those confirmed by both flux \& color metrics, those confirmed with the flux metric only, and those confirmed with the color metric only. In total, 18 targets had confirmed excess in Samples A-E and 44 targets were confirmed with excess in Sample E.} 
\begin{ruledtabular}
\begin{tabular}{ccccc}
 Sample & Observed & Flux \& Color & FluxOnly & ColorOnly \\    \hline
 Samples A-E  & 118  & 4 & 13 & 1\\
 Sample E  & 56 & 22 & 21 & 1
\end{tabular}
\end{ruledtabular}
\end{table}
\endgroup

In Paper I, it was estimated that the false-positive rate for the final 188 infrared excess candidates in Sample E could be as high as 60\% based on the expected frequency of white dwarfs with dust disks or brown dwarf companions. At first glance, our confirmation rate of 79\% for the \textit{Spitzer}-observed Sample E candidates is almost double the expected rate of 40\%. However, recalling that the targets for the Spitzer program were chosen before the Sample criteria in Paper I was established, we can understand this difference as a selection effect. As shown in Table \ref{tab:sampleE-confirmed}, we break down the \textit{Spitzer} confirmation rate in Sample E by the \textit{unWISE} excess designation (color or magnitude), which was a selection factor for our \textit{Spitzer} targets. The \textit{unWISE} magnitude excess in Paper I is analogous to flux excess in this study. As most known white dwarfs with dust disks exhibit both color and flux excess, we have chosen to observe the highest proportion of targets in this category. Candidates with color-only excess as indicated by the \textit{unWISE W1} and \textit{W2} bandpasses are under-observed in comparison. 

Table~\ref{tab:sampleE-confirmed} also shows that only a quarter of the targets in the color-only excess \textit{unWISE} category are confirmed by \textit{Spitzer}. Though only eight color-only \textit{unWISE} excess candidates were observed, these results indicate that color-only excess candidates from Paper I are much more likely to be false positives than candidates with both a color and magnitude excess. Interestingly, there is only one target observed in Sample E with \textit{Spitzer}-confirmed color-only excess, making it considerably rarer than the proportion of candidates indicated by Paper I's \textit{unWISE}-based study. This indicates that the color of contaminants is typically redder than the white dwarf and the confused flux is responsible for the color-only excess determination, consistent with \cite{Barber_2014}. 

Based on the new \textit{Spitzer} data, the 95\% \textit{Spitzer} infrared excess confirmation rate in the Color+Mag \textit{unWISE} excess category indicates that the remaining 9 unobserved targets which have not been previously confirmed are the best candidates for future study. Within the same Color+Mag \textit{unWISE} category, there are 18 known white dwarfs with infrared excess which have all been shown to host circumstellar dust disks. Details of the nine remaining unobserved targets are shown in Table \ref{tab:remaining_targets}. 

\begingroup
\squeezetable
\begin{table}
\caption {\label{tab:sampleE-confirmed} Number of targets in Sample E exhibiting \textit{Spitzer}-confirmed infrared excess with either the color or flux excess metric divided into each \textit{unWISE} excess category as described in Paper I. The published excess column shows the number of known white dwarfs with known excess attributed to dust disks or brown dwarf companions in each category.}
\begin{ruledtabular}
\begin{tabular}{ccccc}
 \textit{unWISE} & Published & Observed & Confirmed & Remaining \\  Excess & Excess & & & \\ \hline
 Color+Mag  & 18 & 20 & 19  & 9 \\
 MagOnly  & 6 & 28 & 23  & 54 \\
 ColorOnly  & 1 & 8 & 2  & 44
\end{tabular}
\end{ruledtabular}
\end{table}
\endgroup

\begingroup
\squeezetable
\begin{table*}[t]
\caption {\label{tab:remaining_targets} Target information for the nine remaining white dwarfs not observed with \textit{Spitzer} from Sample E of Paper I with both color and magnitude excess from \textit{unWISE} photometry. The near-infrared photometric data comes from \textit{VISTA}, \textit{UKIRT}, or \textit{Gemini} and the \textit{WISE} photometry is from \textit{unWISE}. We used the subscript, ``G'', for new Gemini photometry, ``V'' for VISTA photometry, and ``U'' for UKIRT photometry. These white dwarfs constitute the best targets for follow-up observation in the search for infrared excess based on the results of this study.} 
\begin{ruledtabular}
\begin{tabular}{lllcccccc}
 Name & \textit{Gaia} RA & \textit{Gaia} Dec & \multicolumn{5}{c}{$\chi_{\rm{i}}$} & $\Sigma_{\rm{W1-W2}}$ \\
\cline{4-8} 
      & (deg) & (deg)  & \textit{J}  & \textit{H}  & \textit{K}    & \textit{W1} & \textit{W2}                                   \\    \hline
GaiaJ0416+4002 & 64.163225 & 40.042564 &  &  &  & 31.23 & 34.80 & 4.21 \\ 
GaiaJ0508+0535 & 77.059308 & 5.592927 & -0.18$_{\rm{G}}$ & -1.35$_{\rm{G}}$ & -0.11$_{\rm{G}}$ & 8.47 & 10.05 & 3.60 \\ 
GaiaJ0749-3900 & 117.316148 & -39.011862 & -0.25$_{\rm{V}}$ &  & 4.61$_{\rm{V}}$ & 17.23 & 19.53 & 3.73 \\ 
GaiaJ1135-5303 & 173.987460 & -53.055592 & -0.20$_{\rm{V}}$ &  & 3.20$_{\rm{V}}$ & 5.19 & 14.41 & 7.28 \\ 
GaiaJ1412-3546 & 213.242259 & -35.781734 & -0.02$_{\rm{V}}$ &  & 0.67$_{\rm{V}}$ & 12.71 & 20.89 & 7.67 \\ 
GaiaJ1815+1850 & 273.794642 & 18.834127 & 0.23$_{\rm{U}}$ &  &  & 11.51 & 12.58 & 3.70 \\ 
GaiaJ1930-1129 & 292.650443 & -11.496970 & 0.03$_{\rm{V}}$ &  & 4.53$_{\rm{V}}$ & 21.01 & 28.38 & 7.03 \\ 
GaiaJ2004-5127 & 301.223401 & -51.458885 & -0.01$_{\rm{V}}$ &  & 1.92$_{\rm{V}}$ & 6.75 & 10.04 & 4.33 \\ 
GaiaJ2126-2041 & 321.671907 & -20.684547 & -0.25$_{\rm{V}}$ &  & 4.27$_{\rm{V}}$ & 15.58 & 18.39 & 4.63 \\ 
\end{tabular}
\end{ruledtabular}
\end{table*}
\endgroup

\subsection{\textit{Spitzer}-Confirmed Infrared Excesses (Samples A-E)} \label{sec:excess-out-sampleE}
As the targets for follow-up with \textit{Spitzer} were chosen before the criteria for Paper I was finalised, we have also observed 118 white dwarfs with \textit{Spitzer} outside of Sample E. Many of the targets in Samples A-E chosen for follow-up observation with \textit{Spitzer} show evidence of excess in \textit{unWISE} photometry, but were filtered from the final sample due to poor crossmatching, signs of contamination, or other reasons. In Samples A-E, 18 of the 118 \textit{unWISE} white dwarfs were confirmed with \textit{Spitzer} infrared excess. The confirmation rate of 15\% is significantly lower than the 79\% of Sample E candidates. All of the \textit{Spitzer} photometry for observed targets in Samples A-E can be found in the Appendix, Tables \ref{tab:excess_all} and \ref{tab:no-excess_all}. Here, we summarise the white dwarfs with confirmed \textit{Spitzer} infrared excess in Samples A-E. 

Four of the targets show both flux and color excess. One of them, GaiaJ2223-2510, is a heavily polluted DB white dwarf \citep{SALT_2020} previously mis-identified as a hot subdwarf \citep{hot_subdwarf}. Another target, GaiaJ0147+2329, is a known infrared variable with a dusty debris disk also known as Gaia 0145+234 \citep{irvariable}. Its SED is shown in the bottom-right panel of Figure \ref{fig:SED_grid}. GaiaJ1814-7355 and GaiaJ2015+5531 are two new targets with Spitzer-confirmed flux and color excess. Only one target from Samples A-E, GaiaJ1343-0453, exhibits color-only excess.

In Samples A-E, 13 targets exhibit flux-only excess and have existing near-infrared data, many of which were observed with Gemini. Three of these targets, GaiaJ0433+2827, GaiaJ1731-1002, and GaiaJ2026+5925, likely host a stellar or sub-stellar companion based on the strong near-infrared excess in the \textit{J}, \textit{H}, and \textit{K} bands. For the remaining 10, there is no clear evidence of excess in the \textit{J}, \textit{H}, \textit{K} bandpasses, so we are unable to conclude that their excess originates from a binary companion. As with the confirmed infrared excesses from Sample E candidates, these targets are also worthwhile for further investigation into the nature and characteristics of their apparent infrared excess. 

\subsection{Comparison with the Known Sample} \label{sec:compare}
Our \textit{Spitzer} observations confirm a total of 62 white dwarfs with flux or color infrared excess, ten of which have excess likely to be attributed to low-mass companions rather than debris disks based on the observed \textit{J}-band excess. The remaining 52 bright white dwarfs with \textit{Spitzer}-confirmed infrared excess have the potential to double the known sample of white dwarfs with dusty debris disks. Additional spectroscopic studies will allow for further investigations into metal contamination, atmospheric typing of the white dwarf star, and constraining stellar parameters. Among other things, low-resolution follow-up optical spectroscopy can be used to find gas emission lines from dust disks and measure white dwarf atmospheric parameters. High-resolution follow-up can measure atmospheric pollution and infrared spectroscopy will be able to distinguish spectral features of brown dwarf companions from circumstellar dust disks. As \textit{Spitzer} is now decommissioned, this is one of the final large samples of white dwarfs with infrared excess confirmed by the \textit{Spitzer Space Telescope}.

\begin{figure*}[ht!]
\plotone{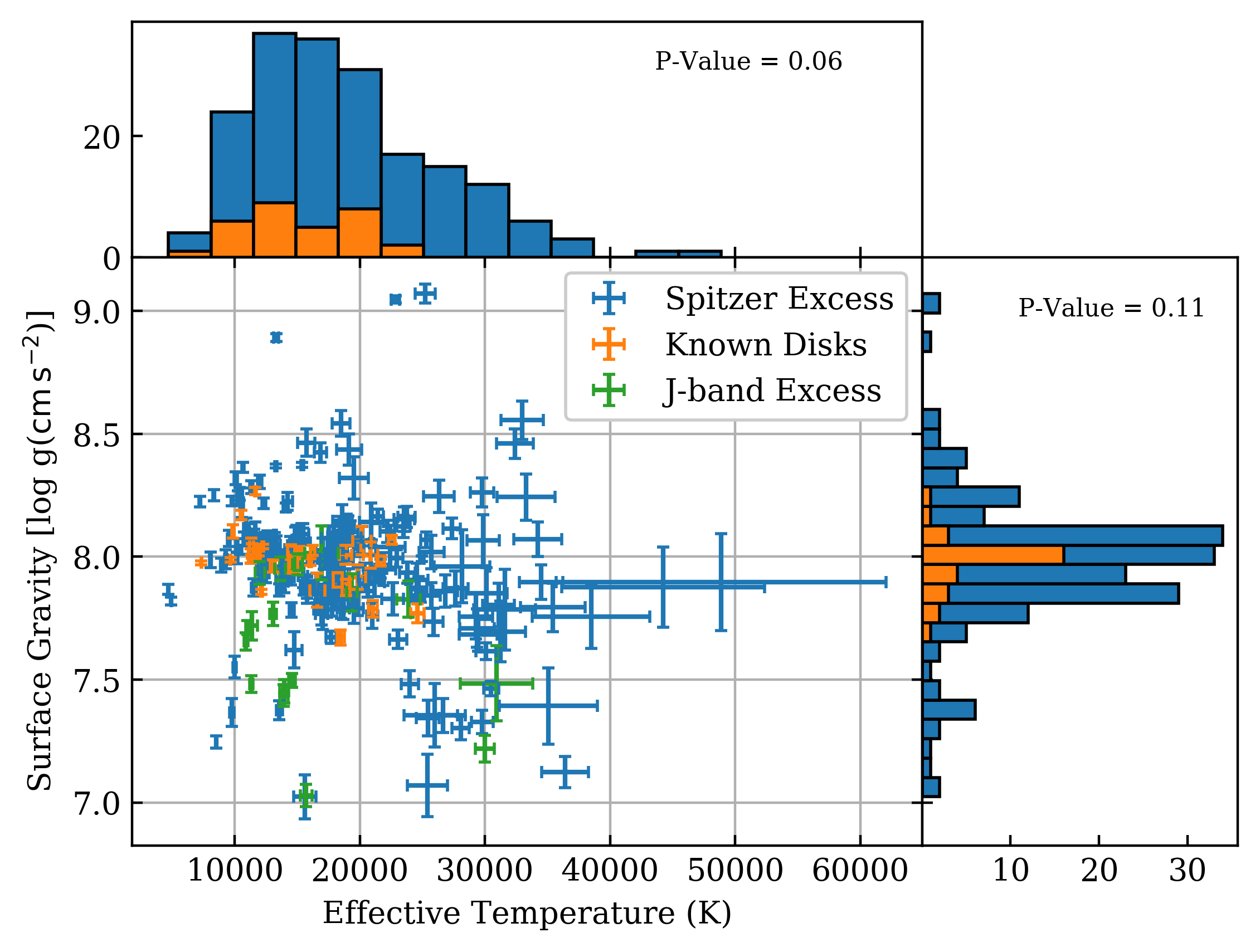}
\caption{Surface gravity and effective temperature distribution of all white dwarfs with confirmed \textit{Spitzer} excess and the known sample of white dwarfs with debris disks. Two-sided Kolmogorov-Smirnov (KS) tests show that the overall distribution of white dwarfs with confirmed excess is not significantly deviating from the distribution of the known sample. \textit{Spitzer}-confirmed white dwarfs with J-band excess are plotted in green, but not included in the distribution. We list the p-values next to each distribution. The white dwarfs with confirmed infrared excess in our study span a wider range in both parameters due to the larger sample size.} \label{fig:cand_known_compare}
\end{figure*}

In Figure \ref{fig:cand_known_compare}, we show a comparison of effective temperature and surface gravity between the known sample of white dwarfs with debris disks and all of the \textit{Spitzer}-confirmed excess white dwarfs. The effective temperature and surface gravity are DA white dwarf model fits to the \textit{Gaia} DR2 photometric data reported in \cite{Gentile_2019WDs}. Two-sided Kolmogorov-Smirnov (KS) tests show that the overall distributions between the two samples are not significantly different, as demonstrated by the high p-values of 0.11 and 0.06 for the surface gravity and effective temperature respectively. Though the distributions are similar, there are candidates outside of the current parameter space occupied by known white dwarf debris disks to the increased sample size. If these \textit{Spitzer} excess white dwarfs are confirmed to host dusty debris disks, they will probe a much wider range in surface gravity and effective temperatures, enabling new discoveries and increasing the range of environments where these disks can exist.

The 62 \textit{Spitzer}-confirmed infrared excess targets presented here are prime targets for further follow-up. For example, \citet{dennihy2020new} and \citet{Melis_2020} independently discovered emission from gaseous debris disks around a total of nine white dwarfs, eight of which were candidates considered in this study. \citet{irvariable} identified GaiaJ0147+2329, shown in Figure \ref{fig:SED_grid}, as a dusty disk with high infrared variability, possibly due to a tidal disruption event in progress. This shows the potential for this list of \textit{Spitzer}-confirmed infrared excess white dwarfs to result in new discoveries.

\section{Conclusion} \label{sec:conclusion}
The first paper in this series, Paper I, identified 188 high-confidence infrared excess candidates constituting the final Sample E \citep{XuCandidates}. In this paper, we discussed the results from photometric observations of 235 white dwarfs using Gemini North's \textit{NIRI} or Gemini South's \textit{F2} and 174 targets with the decommissioned \textit{Spitzer Space Telescope}'s \textit{IRAC}. Infrared photometric measurements were made in the search of excess flux and quantitative measurements of excess were used to evaluate the observed candidates. Most confirmed white dwarfs with dust disks exhibit both flux and color excess. The new observational data found 62 targets with statistically significant infrared excess confirmed by \textit{Spitzer}'s \textit{IRAC} as listed in the Table \ref{tab:excess_all}, 10 of which are likely to be due to stellar or sub-stellar companions. Without additional modelling of the infrared excess, we do not distinguish between stellar or brown dwarf companions in this sample. The remaining 52 bright white dwarfs with infrared excess beyond two microns, 26 of which exhibit both flux and color excess, have the potential to more than double the known sample of white dwarfs with dusty debris disks. With \textit{Spitzer} decommissioned, this study contains one of the final large samples of white dwarfs with infrared excess confirmed by the \textit{Spitzer Space Telescope}.

We caution that without additional spectroscopic observations on the \textit{Spitzer}-confirmed infrared excess targets, it is difficult to determine the source of excess. White dwarfs with low-mass companions and white dwarfs with circumstellar dust disks have similar excess signatures in the infrared. We discussed one way to disentangle these two cases with observations in the near-infrared regime, where unlike dust disks, stellar or sub-stellar companions contribute to the detected excess radiation. We used \textit{J}, \textit{H}, \textit{K} photometry from Gemini North and South Observatory to check for signs of excess characteristic of white dwarf binary systems. However, cooler low-mass companions can not be ruled out in cases with $\chi_{\rm{J}} < 3$, preventing us from determining the occurrence rates of dusty debris disks and low-mass companions around white dwarfs in our sample.

Of the remaining 132 candidates from Sample E of Paper I that were not observed with \textit{Spitzer}, we identify nine high-confidence targets in Table \ref{tab:remaining_targets} for future study based on the 95\% \textit{Spitzer} confirmation rate for \textit{unWISE} excess candidates with both color and magnitude excess.

\section*{Acknowledgements}
We thank the anonymous referee for the constructive comments and suggestions that have improved this manuscript.

This work is based on observations obtained at the international Gemini Observatory, a program of NSF's NOIRLab, which is managed by the Association of Universities for Research in Astronomy (AURA) under a cooperative agreement with the National Science Foundation on behalf of the Gemini Observatory partnership: the National Science Foundation (United States), National Research Council (Canada), Agencia Nacional de Investigaci\'{o}n y Desarrollo (Chile), Ministerio de Ciencia, Tecnolog\'{i}a e Innovaci\'{o}n (Argentina), Minist\'{e}rio da Ciencia, Tecnologia, Inova\c{c}\~{o}es e Comunica\c{c}\~{o}es (Brazil), and Korea Astronomy and Space Science Institute (Republic of Korea). \\
This work is also based in part on observations made with the Spitzer Space Telescope, which is operated by the Jet Propulsion Laboratory, California Institute of Technology under a contract with NASA. Support for this work was provided by NASA through an award issued by JPL/Caltech.\\
ARM acknowledges financial support from the MINECO under the Ram\'on y Cajal program (RYC-2016-20254), the MINECO grant AYA2017-86274-P and the AGAUR grant SGR-661/2017. LKR is grateful to STFC and the Institute of Astronomy, University of Cambridge for funding her PhD studentship.

\vspace{2mm}
\facilities{Gemini:Gillett, Gemini:South, Spitzer}

\software{astropy \citep{astropy},
            DRAGONS \citep{DRAGONS}
          }

\bibliography{sample63}{}

\begin{thebibliography}{}
\expandafter\ifx\csname natexlab\endcsname\relax\def\natexlab#1{#1}\fi
\providecommand{\url}[1]{\href{#1}{#1}}
\providecommand{\dodoi}[1]{doi:~\href{http://doi.org/#1}{\nolinkurl{#1}}}
\providecommand{\doeprint}[1]{\href{http://ascl.net/#1}{\nolinkurl{http://ascl.net/#1}}}
\providecommand{\doarXiv}[1]{\href{https://arxiv.org/abs/#1}{\nolinkurl{https://arxiv.org/abs/#1}}}

\bibitem[{Ahn {et~al.}(2014)Ahn, Alexandroff, Prieto, Anders, Anderson,
  Anderton, Andrews, Aubourg, Bailey, Bastien, Bautista, Beers, Beifiori,
  Bender, Berlind, Beutler, Bhardwaj, Bird, Bizyaev, Blake, Blanton, Blomqvist,
  Bochanski, Bolton, Borde, Bovy, Bradley, Brandt, Brauer, Brinkmann,
  Brownstein, Busca, Carithers, Carlberg, Carnero, Carr, Chiappini, Chojnowski,
  Chuang, Comparat, Crepp, Cristiani, Croft, Cuesta, Cunha, da~Costa, Dawson,
  Lee, Dean, Delubac, Deshpande, Dhital, Ealet, Ebelke, Edmondson, Eisenstein,
  Epstein, Escoffier, Esposito, Evans, Fabbian, Fan, Favole, Castell{\'{a}},
  Alvar, Feuillet, Ak, Finley, Fleming, Font-Ribera, Frinchaboy,
  Galbraith-Frew, Garc{\'{\i}}a-Hern{\'{a}}ndez, P{\'{e}}rez, Ge,
  G{\'{e}}nova-Santos, Gillespie, Girardi, Hern{\'{a}}ndez, Gott, Gunn, Guo,
  Halverson, Harding, Harris, Hasselquist, Hawley, Hayden, Hearty, Dav{\'{o}},
  Ho, Hogg, Holtzman, Honscheid, Huehnerhoff, Ivans, Jackson, Jiang, Johnson,
  Kinemuchi, Kirkby, Klaene, Kneib, Koesterke, Lan, Lang, Goff, Leauthaud, Lee,
  Lee, Long, Loomis, Lucatello, Lupton, Ma, Mack, Mahadevan, Maia, Majewski,
  Malanushenko, Malanushenko, Manchado, Manera, Maraston, Margala, Martell,
  Masters, McBride, McGreer, McMahon, M{\'{e}}nard, M{\'{e}}sz{\'{a}}ros,
  Miralda-Escud{\'{e}}, Miyatake, Montero-Dorta, Montesano, More, Morrison,
  Muna, Munn, Myers, Nguyen, Nichol, Nidever, Noterdaeme, Nuza,
  O{\textquotesingle}Connell, O{\textquotesingle}Connell,
  O{\textquotesingle}Connell, Olmstead, Oravetz, Owen, Padmanabhan,
  Palanque-Delabrouille, Pan, Parejko, Parihar, P{\^{a}}ris, Pepper, Percival,
  P{\'{e}}rez-R{\`{a}}fols, Perottoni, Petitjean, Pieri, Pinsonneault, Prada,
  Price-Whelan, Raddick, Rahman, Rebolo, Reid, Richards, Riffel, Robin,
  Rocha-Pinto, Rockosi, Roe, Ross, Ross, Rossi, Roy, Rubi{\~{n}}o-Martin,
  Sabiu, S{\'{a}}nchez, Santiago, Sayres, Schiavon, Schlegel, Schlesinger,
  Schmidt, Schneider, Schultheis, Sellgren, Seo, Shen, Shetrone, Shu, Simmons,
  Skrutskie, Slosar, Smith, Snedden, Sobeck, Sobreira, Stassun, Steinmetz,
  Strauss, Streblyanska, Suzuki, Swanson, Terrien, Thakar, Thomas, Thompson,
  Tinker, Tojeiro, Troup, Vandenberg, Maga{\~{n}}a, Viel, Vogt, Wake, Weaver,
  Weinberg, Weiner, White, White, Wilson, Wisniewski, Wood-Vasey, Y{\`{e}}che,
  York, Zamora, Zasowski, Zehavi, Zhao, Zheng, \& Zhu}]{SDSS}
Ahn, C.~P., Alexandroff, R., Prieto, C.~A., {et~al.} 2014, The Astrophysical
  Journal Supplement Series, 211, 17, \dodoi{10.1088/0067-0049/211/2/17}

\bibitem[{{Astropy Collaboration} {et~al.}(2013){Astropy Collaboration},
  {Robitaille}, {Tollerud}, {Greenfield}, {Droettboom}, {Bray}, {Aldcroft},
  {Davis}, {Ginsburg}, {Price-Whelan}, {Kerzendorf}, {Conley}, {Crighton},
  {Barbary}, {Muna}, {Ferguson}, {Grollier}, {Parikh}, {Nair}, {Unther},
  {Deil}, {Woillez}, {Conseil}, {Kramer}, {Turner}, {Singer}, {Fox}, {Weaver},
  {Zabalza}, {Edwards}, {Azalee Bostroem}, {Burke}, {Casey}, {Crawford},
  {Dencheva}, {Ely}, {Jenness}, {Labrie}, {Lim}, {Pierfederici}, {Pontzen},
  {Ptak}, {Refsdal}, {Servillat}, \& {Streicher}}]{astropy}
{Astropy Collaboration}, {Robitaille}, T.~P., {Tollerud}, E.~J., {et~al.} 2013,
  \aap, 558, A33, \dodoi{10.1051/0004-6361/201322068}

\bibitem[{Barber {et~al.}(2014)Barber, Kilic, Brown, \&
  Gianninas}]{Barber_2014}
Barber, S.~D., Kilic, M., Brown, W.~R., \& Gianninas, A. 2014, The
  Astrophysical Journal, 786, 77, \dodoi{10.1088/0004-637x/786/2/77}

\bibitem[{{Bergfors} {et~al.}(2014){Bergfors}, {Farihi}, {Dufour}, \&
  {Rocchetto}}]{cool_excess_2014}
{Bergfors}, C., {Farihi}, J., {Dufour}, P., \& {Rocchetto}, M. 2014, \mnras,
  444, 2147, \dodoi{10.1093/mnras/stu1565}

\bibitem[{Bergin {et~al.}(2015)Bergin, Blake, Ciesla, Hirschmann, \&
  Li}]{Bergin2015}
Bergin, E.~A., Blake, G.~A., Ciesla, F., Hirschmann, M.~M., \& Li, J. 2015,
  Proceedings of the National Academy of Sciences, 112, 8965,
  \dodoi{10.1073/pnas.1500954112}

\bibitem[{Brown {et~al.}(2018)Brown, Vallenari, Prusti, de~Bruijne, Babusiaux,
  Bailer-Jones, Biermann, Evans, Eyer, \& et~al.}]{2018Gaia}
Brown, A. G.~A., Vallenari, A., Prusti, T., {et~al.} 2018, Astronomy \&
  Astrophysics, 616, A1, \dodoi{10.1051/0004-6361/201833051}

\bibitem[{{Casewell} {et~al.}(2018){Casewell}, {Littlefair}, {Parsons},
  {Marsh}, {Fortney}, \& {Marley}}]{Casewell_2018}
{Casewell}, S.~L., {Littlefair}, S.~P., {Parsons}, S.~G., {et~al.} 2018,
  \mnras, 481, 5216, \dodoi{10.1093/mnras/sty2599}

\bibitem[{Chambers {et~al.}(2019)Chambers, Magnier, Metcalfe, Flewelling,
  Huber, Waters, Denneau, Draper, Farrow, Finkbeiner, Holmberg, Koppenhoefer,
  Price, Rest, Saglia, Schlafly, Smartt, Sweeney, Wainscoat, Burgett, Chastel,
  Grav, Heasley, Hodapp, Jedicke, Kaiser, Kudritzki, Luppino, Lupton, Monet,
  Morgan, Onaka, Shiao, Stubbs, Tonry, White, Bañados, Bell, Bender, Bernard,
  Boegner, Boffi, Botticella, Calamida, Casertano, Chen, Chen, Cole, Deacon,
  Frenk, Fitzsimmons, Gezari, Gibbs, Goessl, Goggia, Gourgue, Goldman, Grant,
  Grebel, Hambly, Hasinger, Heavens, Heckman, Henderson, Henning, Holman, Hopp,
  Ip, Isani, Jackson, Keyes, Koekemoer, Kotak, Le, Liska, Long, Lucey, Liu,
  Martin, Masci, McLean, Mindel, Misra, Morganson, Murphy, Obaika, Narayan,
  Nieto-Santisteban, Norberg, Peacock, Pier, Postman, Primak, Rae, Rai, Riess,
  Riffeser, Rix, Röser, Russel, Rutz, Schilbach, Schultz, Scolnic, Strolger,
  Szalay, Seitz, Small, Smith, Soderblom, Taylor, Thomson, Taylor, Thakar,
  Thiel, Thilker, Unger, Urata, Valenti, Wagner, Walder, Walter, Watters,
  Werner, Wood-Vasey, \& Wyse}]{panstarrs}
Chambers, K.~C., Magnier, E.~A., Metcalfe, N., {et~al.} 2019, The Pan-STARRS1
  Surveys.
\newblock \doarXiv{1612.05560}

\bibitem[{{Cross, N. J. G.} {et~al.}(2012){Cross, N. J. G.}, {Collins, R. S.},
  {Mann, R. G.}, {Read, M. A.}, {Sutorius, E. T. W.}, {Blake, R. P.},
  {Holliman, M.}, {Hambly, N. C.}, {Emerson, J. P.}, {Lawrence, A.}, \&
  {Noddle, K. T.}}]{VISTA_Archive}
{Cross, N. J. G.}, {Collins, R. S.}, {Mann, R. G.}, {et~al.} 2012, A\&A, 548,
  A119, \dodoi{10.1051/0004-6361/201219505}

\bibitem[{{Cutri} {et~al.}(2013){Cutri}, {Wright}, {Conrow}, {Fowler},
  {Eisenhardt}, {Grillmair}, {Kirkpatrick}, {Masci}, {McCallon}, {Wheelock},
  {Fajardo-Acosta}, {Yan}, {Benford}, {Harbut}, {Jarrett}, {Lake}, {Leisawitz},
  {Ressler}, {Stanford}, {Tsai}, {Liu}, {Helou}, {Mainzer}, {Gettings},
  {Gonzalez}, {Hoffman}, {Marsh}, {Padgett}, {Skrutskie}, {Beck}, {Papin}, \&
  {Wittman}}]{ALLWISE}
{Cutri}, R.~M., {Wright}, E.~L., {Conrow}, T., {et~al.} 2013, {Explanatory
  Supplement to the AllWISE Data Release Products}, Explanatory Supplement to
  the AllWISE Data Release Products

\bibitem[{Debes {et~al.}(2011)Debes, Hoard, Wachter, Leisawitz, \&
  Cohen}]{WIREDII}
Debes, J.~H., Hoard, D.~W., Wachter, S., Leisawitz, D.~T., \& Cohen, M. 2011,
  The Astrophysical Journal Supplement Series, 197, 38,
  \dodoi{10.1088/0067-0049/197/2/38}

\bibitem[{Debes \& Sigurdsson(2002)}]{Debes2002}
Debes, J.~H., \& Sigurdsson, S. 2002, The Astrophysical Journal, 572, 556,
  \dodoi{10.1086/340291}

\bibitem[{Dennihy {et~al.}(2020{\natexlab{a}})Dennihy, Farihi, Fusillo, \&
  Debes}]{Dennihy_2020confusion}
Dennihy, E., Farihi, J., Fusillo, N. P.~G., \& Debes, J.~H. 2020{\natexlab{a}},
  The Astrophysical Journal, 891, 97, \dodoi{10.3847/1538-4357/ab7249}

\bibitem[{Dennihy {et~al.}(2020{\natexlab{b}})Dennihy, Xu, Lai, Bonsor,
  Clemens, Dufour, Gänsicke, Fusillo, Hardy, Hegedus, Hermes, Kaiser,
  Kissler-Patig, Klein, Manser, \& Reding}]{dennihy2020new}
Dennihy, E., Xu, S., Lai, S., {et~al.} 2020{\natexlab{b}}, The Astrophysical
  Journal, 905, 5, \dodoi{10.3847/1538-4357/abc339}

\bibitem[{Dye {et~al.}(2017)Dye, Lawrence, Read, Fan, Kerr, Varricatt, Furnell,
  Edge, Irwin, Hambly, \& et~al.}]{UHS_DR1}
Dye, S., Lawrence, A., Read, M.~A., {et~al.} 2017, Monthly Notices of the Royal
  Astronomical Society, 473, 5113–5125, \dodoi{10.1093/mnras/stx2622}

\bibitem[{Eikenberry {et~al.}(2012)Eikenberry, Bandyopadhyay, Bennett, Bessoff,
  Branch, Charcos, Corley, Dewitt, Eriksen, Elston, Frommeyer, Gonzalez, Hanna,
  Herlevich, Hon, Julian, Julian, Lasso, Marin-Franch, Marti, Murphey, Raines,
  Rambold, Rashkind, Warner, Leckie, Gardhouse, Fletcher, Hardy, Dunn, Wooff,
  \& Pazder}]{F2}
Eikenberry, S., Bandyopadhyay, R., Bennett, J.~G., {et~al.} 2012, in
  Ground-based and Airborne Instrumentation for Astronomy IV, ed. I.~S. McLean,
  S.~K. Ramsay, \& H.~Takami, Vol. 8446, International Society for Optics and
  Photonics (SPIE), 136 -- 148.
\newblock \url{https://doi.org/10.1117/12.925679}

\bibitem[{{Farihi}(2016)}]{Farihi2016}
{Farihi}, J. 2016, \nar, 71, 9, \dodoi{10.1016/j.newar.2016.03.001}

\bibitem[{Farihi \& Christopher(2004)}]{BD1}
Farihi, J., \& Christopher, M. 2004, The Astronomical Journal, 128,
  1868–1871, \dodoi{10.1086/423919}

\bibitem[{Farihi {et~al.}(2008)Farihi, Zuckerman, \& Becklin}]{Farihi_2008}
Farihi, J., Zuckerman, B., \& Becklin, E.~E. 2008, The Astrophysical Journal,
  674, 431, \dodoi{10.1086/521715}

\bibitem[{{Geier} {et~al.}(2017){Geier}, {{\O}stensen}, {Nemeth}, {Gentile
  Fusillo}, {G{\"a}nsicke}, {Telting}, {Green}, \&
  {Schaffenroth}}]{hot_subdwarf}
{Geier}, S., {{\O}stensen}, R.~H., {Nemeth}, P., {et~al.} 2017, \aap, 600, A50,
  \dodoi{10.1051/0004-6361/201630135}

\bibitem[{{Gentile Fusillo} {et~al.}(2019){Gentile Fusillo}, {Tremblay},
  {G{\"a}nsicke}, {Manser}, {Cunningham}, {Cukanovaite}, {Hollands}, {Marsh},
  {Raddi}, {Jordan}, {Toonen}, {Geier}, {Barstow}, \&
  {Cummings}}]{Gentile_2019WDs}
{Gentile Fusillo}, N.~P., {Tremblay}, P.-E., {G{\"a}nsicke}, B.~T., {et~al.}
  2019, \mnras, 482, 4570, \dodoi{10.1093/mnras/sty3016}

\bibitem[{{Girven} {et~al.}(2011){Girven}, {G{\"a}nsicke}, {Steeghs}, \&
  {Koester}}]{Girven2011MNRAS.417.1210G}
{Girven}, J., {G{\"a}nsicke}, B.~T., {Steeghs}, D., \& {Koester}, D. 2011,
  \mnras, 417, 1210, \dodoi{10.1111/j.1365-2966.2011.19337.x}

\bibitem[{{Gonz{\'a}lez-Fern{\'a}ndez}
  {et~al.}(2018){Gonz{\'a}lez-Fern{\'a}ndez}, {Hodgkin}, {Irwin},
  {Gonz{\'a}lez-Solares}, {Koposov}, {Lewis}, {Emerson}, {Hewett},
  {Yolda{\textcommabelow s}}, \& {Riello}}]{vista-mko-conversion}
{Gonz{\'a}lez-Fern{\'a}ndez}, C., {Hodgkin}, S.~T., {Irwin}, M.~J., {et~al.}
  2018, \mnras, 474, 5459, \dodoi{10.1093/mnras/stx3073}

\bibitem[{Hambly {et~al.}(2008)Hambly, Collins, Cross, Mann, Read, Sutorius,
  Bond, Bryant, Emerson, Lawrence, Rimoldini, Stewart, Williams, Adamson,
  Hirst, Dye, \& Warren}]{WFCAM_Archive}
Hambly, N.~C., Collins, R.~S., Cross, N. J.~G., {et~al.} 2008, Monthly Notices
  of the Royal Astronomical Society, 384, 637,
  \dodoi{10.1111/j.1365-2966.2007.12700.x}

\bibitem[{Harrison {et~al.}(2018)Harrison, Bonsor, \&
  Madhusudhan}]{Harrison_2018}
Harrison, J. H.~D., Bonsor, A., \& Madhusudhan, N. 2018, Monthly Notices of the
  Royal Astronomical Society, 479, 3814–3841, \dodoi{10.1093/mnras/sty1700}

\bibitem[{Hodapp {et~al.}(2000)Hodapp, Hora, Graves, Irwin, Yamada, Douglass,
  Young, \& Robertson}]{NIRI}
Hodapp, K.-W., Hora, J., Graves, E., {et~al.} 2000, Proc SPIE, 4008, 1334

\bibitem[{Hodgkin {et~al.}(2009)Hodgkin, Irwin, Hewett, \&
  Warren}]{2mass-mko-conversion}
Hodgkin, S., Irwin, M., Hewett, P., \& Warren, S. 2009, Mon. Not. Roy. Astron.
  Soc., 394, 675, \dodoi{10.1111/j.1365-2966.2008.14387.x}

\bibitem[{{Holberg} \& {Bergeron}(2006)}]{Bergeron_model}
{Holberg}, J.~B., \& {Bergeron}, P. 2006, \aj, 132, 1221,
  \dodoi{10.1086/505938}

\bibitem[{Jeffery {et~al.}(2020)Jeffery, Miszalski, \& Snowdon}]{SALT_2020}
Jeffery, C.~S., Miszalski, B., \& Snowdon, E. 2020, Monthly Notices of the
  Royal Astronomical Society, 501, 623–642, \dodoi{10.1093/mnras/staa3648}

\bibitem[{{Jura}(2003)}]{Jura2003}
{Jura}, M. 2003, \apjl, 584, L91, \dodoi{10.1086/374036}

\bibitem[{Jura \& Young(2014)}]{Jura_Extrasolar_Cosmochemistry}
Jura, M., \& Young, E. 2014, Annual Review of Earth and Planetary Sciences, 42,
  45, \dodoi{10.1146/annurev-earth-060313-054740}

\bibitem[{{Labrie} {et~al.}(2019){Labrie}, {Anderson}, {C{\'a}rdenes},
  {Simpson}, \& {Turner}}]{DRAGONS}
{Labrie}, K., {Anderson}, K., {C{\'a}rdenes}, R., {Simpson}, C., \& {Turner},
  J. E.~H. 2019, Astronomical Society of the Pacific Conference Series, Vol.
  523, {DRAGONS - Data Reduction for Astronomy from Gemini Observatory North
  and South}, ed. P.~J. {Teuben}, M.~W. {Pound}, B.~A. {Thomas}, \& E.~M.
  {Warner}, 321

\bibitem[{Lawrence {et~al.}(2007)Lawrence, Warren, Almaini, Edge, Hambly,
  Jameson, Lucas, Casali, Adamson, Dye, Emerson, Foucaud, Hewett, Hirst,
  Hodgkin, Irwin, Lodieu, McMahon, Simpson, Smail, Mortlock, \&
  Folger}]{UKIDSS}
Lawrence, A., Warren, S.~J., Almaini, O., {et~al.} 2007, Monthly Notices of the
  Royal Astronomical Society, 379, 1599,
  \dodoi{10.1111/j.1365-2966.2007.12040.x}

\bibitem[{{Leggett} {et~al.}(2015){Leggett}, {Morley}, {Marley}, \&
  {Saumon}}]{sandy-nir-dwarfs}
{Leggett}, S.~K., {Morley}, C.~V., {Marley}, M.~S., \& {Saumon}, D. 2015, \apj,
  799, 37, \dodoi{10.1088/0004-637X/799/1/37}

\bibitem[{Longstaff {et~al.}(2019)Longstaff, Casewell, Wynn, Page, Williams,
  Braker, \& Maxted}]{Longstaff_2019}
Longstaff, E.~S., Casewell, S.~L., Wynn, G.~A., {et~al.} 2019, Monthly Notices
  of the Royal Astronomical Society, 484, 2566, \dodoi{10.1093/mnras/stz127}

\bibitem[{Malamud {et~al.}(2020)Malamud, Grishin, \& Brouwers}]{Malamud_2020}
Malamud, U., Grishin, E., \& Brouwers, M. 2020, Monthly Notices of the Royal
  Astronomical Society, 501, 3806, \dodoi{10.1093/mnras/staa3940}

\bibitem[{Malamud \& Perets(2020)}]{Malamud_2020(2)}
Malamud, U., \& Perets, H.~B. 2020, Monthly Notices of the Royal Astronomical
  Society, 492, 5561–5581, \dodoi{10.1093/mnras/staa142}

\bibitem[{Maxted {et~al.}(2006)Maxted, Napiwotzki, Dobbie, \& Burleigh}]{BD2}
Maxted, P. F.~L., Napiwotzki, R., Dobbie, P.~D., \& Burleigh, M.~R. 2006,
  Nature, 442, 543, \dodoi{10.1038/nature04987}

\bibitem[{Melis {et~al.}(2020)Melis, Klein, Doyle, Weinberger, Zuckerman, \&
  Dufour}]{Melis_2020}
Melis, C., Klein, B., Doyle, A.~E., {et~al.} 2020, The Astrophysical Journal,
  905, 56, \dodoi{10.3847/1538-4357/abbdfa}

\bibitem[{Rafikov \& Garmilla(2012)}]{Rafikov_2012}
Rafikov, R.~R., \& Garmilla, J.~A. 2012, The Astrophysical Journal, 760, 123,
  \dodoi{10.1088/0004-637x/760/2/123}

\bibitem[{Rappaport {et~al.}(2017)Rappaport, Vanderburg, Nelson, Gary, Kaye,
  Kalomeni, Howell, Thorstensen, Lachapelle, Lundy, \&
  St-Antoine}]{Rappaport_2017}
Rappaport, S., Vanderburg, A., Nelson, L., {et~al.} 2017, Monthly Notices of
  the Royal Astronomical Society, 471, 948, \dodoi{10.1093/mnras/stx1611}

\bibitem[{{Rebassa-Mansergas} {et~al.}(2016){Rebassa-Mansergas}, {Ren},
  {Parsons}, {G{\"a}nsicke}, {Schreiber}, {Garc{\'\i}a-Berro}, {Liu}, \&
  {Koester}}]{2016Rebassa_WDM}
{Rebassa-Mansergas}, A., {Ren}, J.~J., {Parsons}, S.~G., {et~al.} 2016, \mnras,
  458, 3808, \dodoi{10.1093/mnras/stw554}

\bibitem[{{Rebassa-Mansergas} {et~al.}(2019){Rebassa-Mansergas}, {Solano},
  {Xu}, {Rodrigo}, {Jim{\'e}nez-Esteban}, \& {Torres}}]{Rebassa_Gaia}
{Rebassa-Mansergas}, A., {Solano}, E., {Xu}, S., {et~al.} 2019, \mnras, 489,
  3990, \dodoi{10.1093/mnras/stz2423}

\bibitem[{{Rocchetto} {et~al.}(2015){Rocchetto}, {Farihi}, {G{\"a}nsicke}, \&
  {Bergfors}}]{Rocchetto2015MNRAS.449..574R}
{Rocchetto}, M., {Farihi}, J., {G{\"a}nsicke}, B.~T., \& {Bergfors}, C. 2015,
  \mnras, 449, 574, \dodoi{10.1093/mnras/stv282}

\bibitem[{{Schlafly} {et~al.}(2019){Schlafly}, {Meisner}, \&
  {Green}}]{unWISE2019}
{Schlafly}, E.~F., {Meisner}, A.~M., \& {Green}, G.~M. 2019, \apjs, 240, 30,
  \dodoi{10.3847/1538-4365/aafbea}

\bibitem[{Skrutskie {et~al.}(2006)Skrutskie, Cutri, Stiening, Weinberg,
  Schneider, Carpenter, Beichman, Capps, Chester, Elias, Huchra, Liebert,
  Lonsdale, Monet, Price, Seitzer, Jarrett, Kirkpatrick, Gizis, Howard, Evans,
  Fowler, Fullmer, Hurt, Light, Kopan, Marsh, McCallon, Tam, Dyk, \&
  Wheelock}]{2MASS}
Skrutskie, M.~F., Cutri, R.~M., Stiening, R., {et~al.} 2006, The Astronomical
  Journal, 131, 1163, \dodoi{10.1086/498708}

\bibitem[{{Steele} {et~al.}(2011){Steele}, {Burleigh}, {Dobbie}, {Jameson},
  {Barstow}, \& {Satterthwaite}}]{Steele2011MNRAS.416.2768S}
{Steele}, P.~R., {Burleigh}, M.~R., {Dobbie}, P.~D., {et~al.} 2011, \mnras,
  416, 2768, \dodoi{10.1111/j.1365-2966.2011.19225.x}

\bibitem[{Steele {et~al.}(2013)Steele, Saglia, Burleigh, Marsh, Gänsicke,
  Lawrie, Cappetta, Girven, \& Napiwotzki}]{BD3}
Steele, P.~R., Saglia, R.~P., Burleigh, M.~R., {et~al.} 2013, Monthly Notices
  of the Royal Astronomical Society, 429, 3492, \dodoi{10.1093/mnras/sts620}

\bibitem[{{Wang} {et~al.}(2019){Wang}, {Jiang}, {Ge}, {Cutri}, {Jiang},
  {Sheng}, {Zhou}, {Bauer}, {Mainzer}, \& {Wright}}]{irvariable}
{Wang}, T.-G., {Jiang}, N., {Ge}, J., {et~al.} 2019, arXiv e-prints,
  arXiv:1910.04314.
\newblock \doarXiv{1910.04314}

\bibitem[{{Werner} {et~al.}(2004){Werner}, {Roellig}, {Low}, {Rieke}, {Rieke},
  {Hoffmann}, {Young}, {Houck}, {Brandl}, {Fazio}, {Hora}, {Gehrz}, {Helou},
  {Soifer}, {Stauffer}, {Keene}, {Eisenhardt}, {Gallagher}, {Gautier}, {Irace},
  {Lawrence}, {Simmons}, {Van Cleve}, {Jura}, {Wright}, \&
  {Cruikshank}}]{Spitzer_orig}
{Werner}, M.~W., {Roellig}, T.~L., {Low}, F.~J., {et~al.} 2004, \apjs, 154, 1,
  \dodoi{10.1086/422992}

\bibitem[{{Wilson} {et~al.}(2019){Wilson}, {Farihi}, {G{\"a}nsicke}, \&
  {Swan}}]{wilson-unbiased-freq}
{Wilson}, T.~G., {Farihi}, J., {G{\"a}nsicke}, B.~T., \& {Swan}, A. 2019,
  \mnras, 487, 133, \dodoi{10.1093/mnras/stz1050}

\bibitem[{Xu {et~al.}(2020)Xu, Lai, \& Dennihy}]{XuCandidates}
Xu, S., Lai, S., \& Dennihy, E. 2020, The Astrophysical Journal, 902, 127,
  \dodoi{10.3847/1538-4357/abb3fc}

\bibitem[{Xu {et~al.}(2021)Xu, Lai, \& Dennihy}]{Xu_2021}
---. 2021, The Astrophysical Journal, 916, 121,
  \dodoi{10.3847/1538-4357/ac135b}

\end{thebibliography}

\appendix
\setcounter{table}{0}
\renewcommand{\thetable}{A\arabic{table}}
\renewcommand*{\theHtable}{A\arabic{table}} 
\setcounter{figure}{0}  
\renewcommand\thefigure{A\arabic{figure}} 
\renewcommand*{\theHfigure}{A\arabic{figure}}
\section{Gemini Observatory Near Infrared Imaging Diagnostics} \label{Appendix}

\subsection{Filter Transformations} \label{appendix:filter_transformations}
In this work, all of the magnitudes are reported in the \textit{Mauna Kea Observatory} (\textit{MKO}) system and the appropriate filter transformations were applied. As not all fields have \textit{MKO}-system photometry available, we have used \textit{2MASS} as the universal calibrator. For our standard stars, we converted the \textit{2MASS} \textit{J}, \textit{H}, \textit{$K_{\rm{s}}$} magnitudes into the \textit{MKO} system using the transformations which were measured empirically for regions of low reddening, specifically Equations 6, 7, and 8 from \cite{2mass-mko-conversion}.

For \textit{NIRI}, all three of its near-infrared filters are similar to \textit{UKIDSS}, which is already in the \textit{MKO} system. We transformed the \textit{2MASS} photometry of reference stars according to the above color equations before measuring the target magnitudes. For \textit{F2}, although \textit{J} and \textit{H} filters are in the \textit{MKO} system, the \textit{$K_{\rm{s}}$} filter profile is very similar to the \textit{2MASS} \textit{$K_{\rm{s}}$} filter \citep{sandy-nir-dwarfs}. Thus, the transformation of \textit{2MASS} \textit{$K_{\rm{s}}$} to the \textit{MKO} system is applied after measuring the target magnitude.

Comparisons between \textit{2MASS} and \textit{UKIDSS} photometry show a systematic deviation in the \textit{K} filter wavelength regime when the targets are fainter than roughly 15.5 magnitude due to a sensitivity limit in \textit{2MASS} \citep{2MASS}. For this reason, we choose reference stars with \textit{$K_{\rm{s}}$} magnitude brighter than 15.5 mag unless there are fewer than two reference stars within frame. The same reference stars used to perform photometry in the \textit{K}-bandpass are reused for both the \textit{J}- and \textit{H}-bandpasses whenever possible.

\begin{figure}[ht!]
\plotone{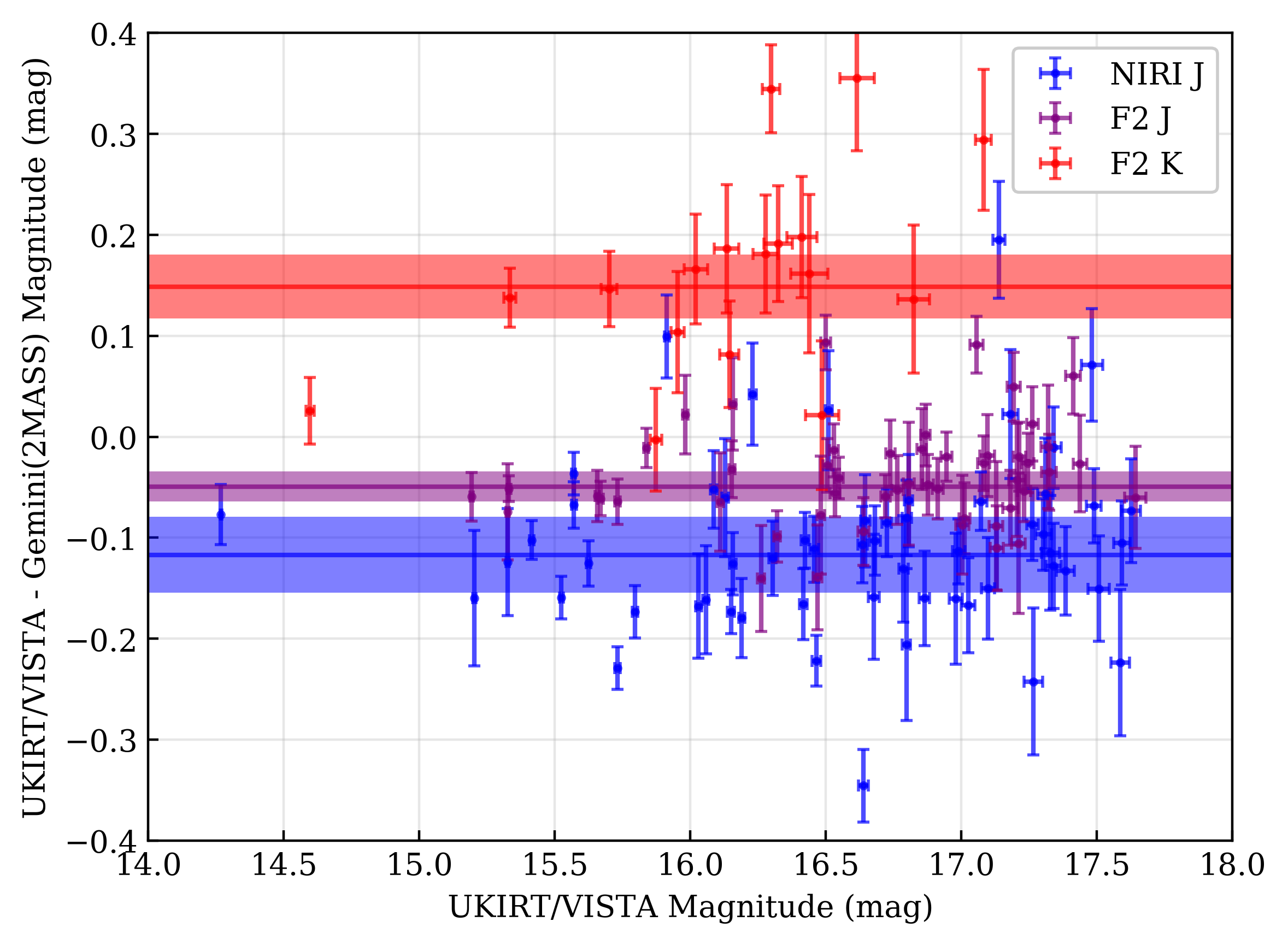}
\caption{Comparison of the \textit{2MASS}-calibrated Gemini photometry against \textit{UKIRT} or \textit{VISTA} survey photometry before linear correction. The applied offsets are shown as the solid horizontal lines and their errors are represented by the shaded regions. We found no offsets for \textit{NIRI H}, \textit{NIRI K}, and \textit{F2 H}. \label{fig:photo-offset}}
\end{figure}

Further comparisons between our \textit{2MASS}-calibrated Gemini photometry against existing near-infrared photometry from \textit{UKIRT}'s \textit{WFCAM} and \textit{VISTA} show some systematic linear offset. Correcting for this offset was found to improve the reliability of the Gemini photometry in fitting with white dwarf photospheric models. We show this offset by comparing the uncorrected Gemini photometry with a sample of white dwarfs that have existing survey photometry (\textit{UKIRT}/\textit{VISTA}) in Figure \ref{fig:photo-offset}. The linear offsets for \textit{NIRI J} and \textit{F2 J} were found to be $-0.12\pm0.04$ mag and $-0.05\pm0.01$ mag respectively when compared against \textit{WFCAM J}. The linear offset for \textit{$K_{\rm{s}}$} was $0.15\pm0.03$ mag when compared against \textit{VISTA $K_{\rm{s}}$}. All offsets were calculated with a weighted average where higher weight was applied for brighter stars with lower uncertainty. The \textit{VISTA} magnitudes were converted into the \textit{MKO} system according to the literature, specifically Equations 16, 17, and 18 from \cite{vista-mko-conversion}. For the remaining bandpasses, \textit{NIRI H}, \textit{NIRI K}, and \textit{F2 H}, there was not enough existing photometry to determine robust statistical offsets and what little data there was did not show significant discrepancy between the Gemini and \textit{WFCAM} or \textit{VISTA} photometry. Therefore, no linear offsets were applied to \textit{NIRI H}, \textit{NIRI K}, and \textit{F2 H}. 

There are also no significant trends in the zeropoint difference with calibrator brightness or color, suggesting the photometric transformation and linearity correction is acceptable. However the systematic offset suggests there are unrecognised errors at the $\lesssim$10\% level  in the \textit{2MASS} to \textit{MKO} photometric transformation and/or in the \textit{NIRI}/\textit{F2} linearity correction (the \textit{2MASS} stars are brighter).  The \textit{2MASS} photometry is more uncertain than the survey photometry, and the linearity correction and system transformation is larger between \textit{2MASS} and raw data, therefore we apply this offset to our measurements to put them onto the \textit{MKO} system.

\subsection{Photometric Data Quality Assessment}
This section discusses a variety of factors which affect photometric data quality and evaluates how each factor affects the result. We will discuss observing conditions, image quality, and misalignment of coordinates in the output image of our observations with both \textit{NIRI} and \textit{F2}.

Of the 16 programmes observed using Gemini's North and South near-infrared imagers, eight of the largest programmes were poor weather programmes, characterised by observing condition constraints of Cloud Cover (CC) in the 70th, 80th, or ``Any'' percentiles. Each percentile corresponds to a percentage of time with a certain transparency based on long term data for Mauna Kea. Higher percentiles indicate the potential for more cloud coverage, and thus a greater loss of signal. Poor weather proposals also do not place any restriction on the desired Image Quality (IQ) and Water Vapour (WV) content. For the 215 targets observed by either Gemini's \textit{NIRI} or \textit{F2} instrument, 150 were observed in poor weather conditions, split between 69\% (80/116) of \textit{NIRI} observations and 88\% (111/126) of \textit{F2} observations. Majority of targets were observed in photometric conditions of IQ85 or better, indicating a FWHM of less than 0.85 arcseconds in the \textit{J} bandpass. 

\begin{figure*}
\gridline{\fig{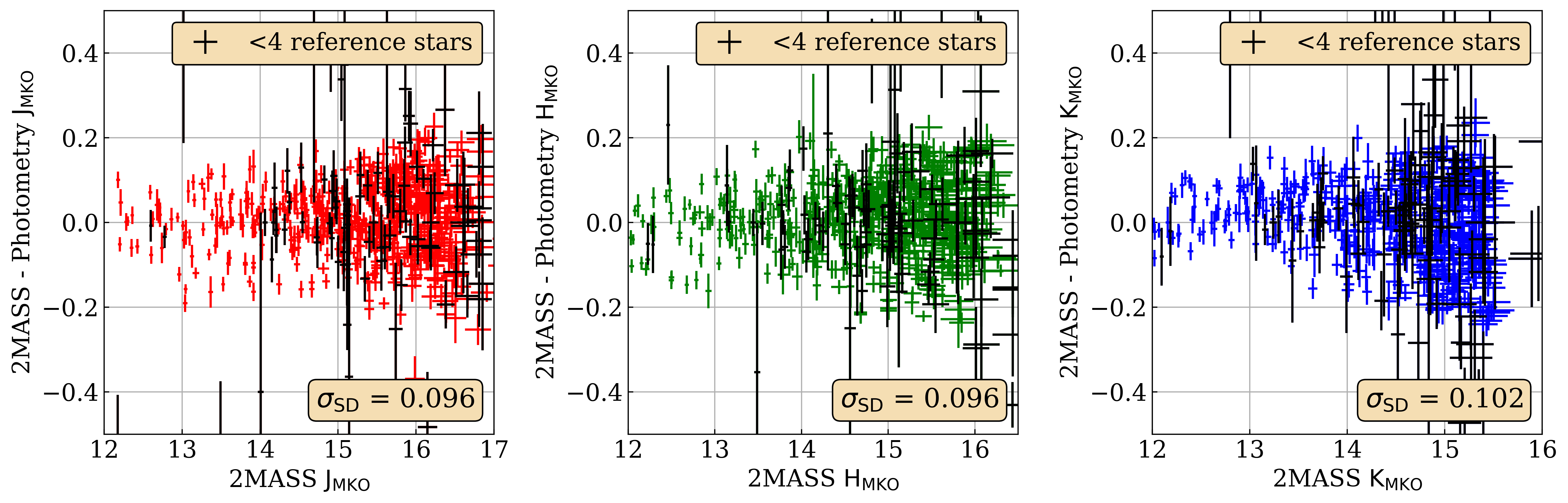}{0.9\textwidth}{(NIRI)}
          }
\gridline{\fig{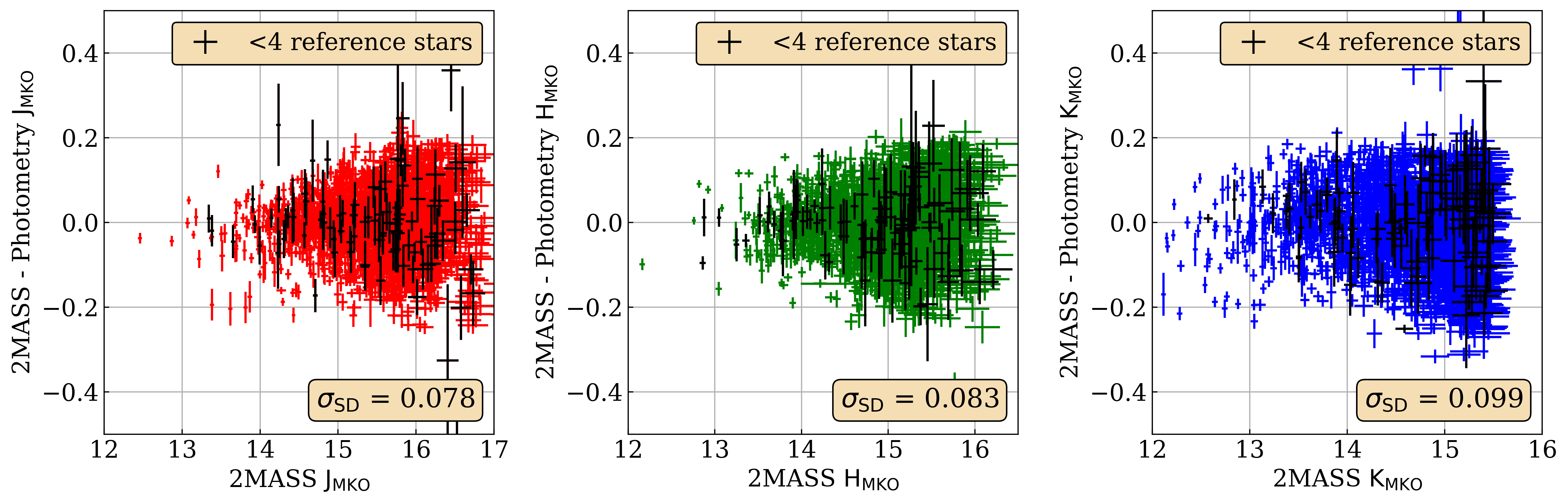}{0.9\textwidth}{(F2)}
          }
\caption{Gemini near-infrared photometry compared against \textit{2MASS} \textit{J}, \textit{H}, \textit{K} photometry of reference stars within the observed field of view. All photometry are converted into the \textit{MKO} system for comparison. The standard deviation ($\sigma_{\rm{SD}}$) values are displayed. There is a higher dispersion in the measured magnitudes for targets in fields where there are few reference stars (as shown by the black data points) and this effect is more pronounced for \textit{NIRI} than for \textit{F2} because of the smaller field of view. The number of reference stars used affects the quality of the photometry. \label{fig:2mass-gemini-compare}}
\end{figure*}

In Figure~\ref{fig:2mass-gemini-compare}, all of the measured aperture photometry obtained from our algorithm is compared against the \textit{2MASS} photometry of the reference stars. The figure does not show an appreciable systematic deviation with brightness between our photometry with \textit{2MASS} in the \textit{J}, \textit{H}, \textit{K} filters for either \textit{NIRI} or \textit{F2} instruments. We find that although the \textit{2MASS} precision decreases for fainter stars, the primary contributing factor to the Gemini photometric precision is the number of reference stars used to calculate the Gemini photometry. Furthermore, even though the majority of the targets were observed under poor weather conditions, we find that the standard deviation in the difference between \textit{2MASS} and our photometry to be $\lesssim$0.1 mag. Since the field of view of \textit{F2} is 6'x 6' circular field compared against the 120"x 120" square field of \textit{NIRI}, there are often more reference stars in \textit{F2} which improves both the accuracy and precision of the measured magnitudes. We conclude that the measured Gemini near-infrared photometry is reliable even if targets are observed under poor weather conditions as long as there are a good number of bright \textit{2MASS} reference stars within the field of view.

\begin{figure}[ht!]
\plotone{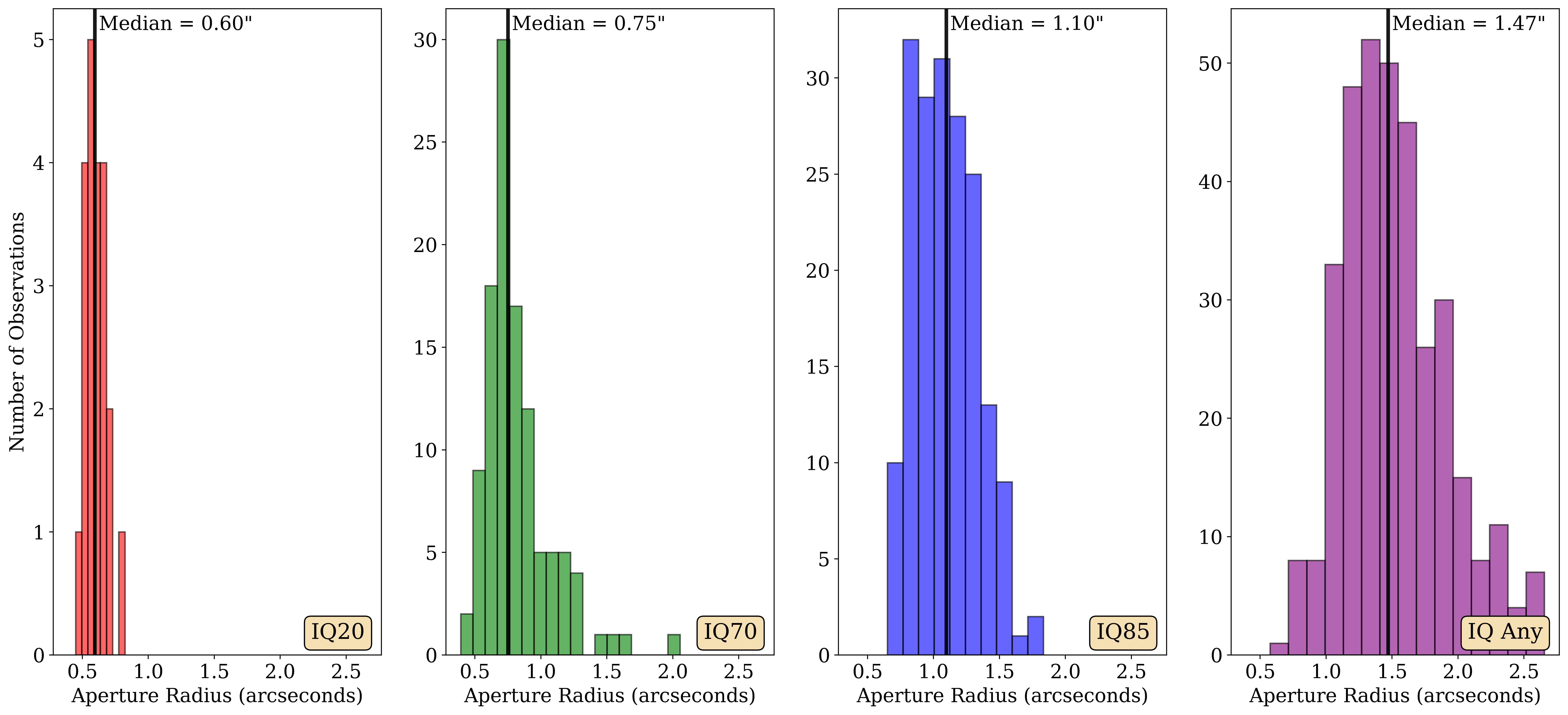}
\caption{Relationship between aperture radius and Image Quality (IQ) based on the combined data from Gemini North's \textit{NIRI} and Gemini South's \textit{F2} in this study. The aperture radius is comparable between instruments and bandpasses, but the same trend can be observed for any individual filter and instrument. The solid black line indicates the median aperture radius for the indicated IQ classification. The aperture radius used in photometry of a target and its reference stars increases with deteriorating IQ. \label{fig:ap_IQ}}
\end{figure}

The Image Quality (IQ) at Gemini Observatory is a percentile defined for a target at zenith with a profile FWHM below a wavelength-dependent maximum threshold. The percentile following the IQ represents the percentage of time when the FWHM is below a defined threshold and is linked to astronomical seeing\footnote{\url{https://www.gemini.edu/observing/telescopes-and-sites/sites/}}. Point sources of the reference stars and our target were modeled using \textit{astropy}'s \textit{photutils}, where the aperture radius was designated as three times the median of the FWHM among all stars detected in the frame determined by the PSF fitting. Figure~\ref{fig:ap_IQ} shows that the median aperture radius in each IQ category increases with the deteriorating quality, independently recovering the desired result without directly referencing the observed IQ of each frame.

The World Coordinates System (WCS) of the output images for both \textit{NIRI} and \textit{F2} can be offset from the true WCS when compared against other public surveys. In all of the observations performed in this study, a unique linear correction applied to each individual image was sufficient to correct for the WCS offset compared against \textit{2MASS}. Some images show non-linear warping, but the effects are negligible compared to the linear offset. We find that the magnitude of the median offset for \textit{NIRI} is typically 1.1", split into -1.06" in the X-direction and -0.33" in the Y-direction, while the median \textit{F2} offset is 8.9", split into 5.47" in the X-direction and 7.07" in the Y-direction. 

We also assess how the photometric zeropoint determined from \textit{2MASS} reference stars is dependent on the observation conditions. Table~\ref{tab:zp_cc} shows the median zeropoints in magnitude for each filter in the two Gemini instruments. For the set of observations performed in this study, the measured zeropoints are roughly comparable between CC50 and CC70, but decrease sharply with CC80, consistent with a greater loss of signal under those observing conditions. 

\begin{deluxetable*}{cccccc}
\tablecaption{Mean photometric magnitude zeropoint relationship with observed cloud cover percentile for \textit{NIRI} and \textit{F2} programmes observed under CC50 to CC80 conditions. The zeropoint of individual frames depends also on the airmass and extinction correction. We have assumed an airmass of 1.0 for these values. \label{tab:zp_cc}}
\tablewidth{0pt}
\tablehead{
\colhead{Instrument \& Filter}  & \colhead{CC50} & \colhead{CC70} & \colhead{CC80} \\
\colhead{}  & \colhead{mag} & \colhead{mag} & \colhead{mag}
}
\startdata
\textit{NIRI J} & 23.66$\,\pm\, 0.03$ & 23.60$\,\pm\, 0.03$ & 23.17$\,\pm\, 0.08$ \\
\textit{NIRI H} & 23.80$\,\pm\, 0.08$ & 23.67$\,\pm\, 0.06$ & 23.25$\,\pm\, 0.10$ \\
\textit{NIRI K} & 23.12$\,\pm\, 0.04$ & 22.86$\,\pm\, 0.22$ & 22.58$\,\pm\, 0.11$ \\
\textit{F2 J} & 24.84$\,\pm\, 0.03$ & 24.89$\,\pm\, 0.02$ & 24.71$\,\pm\, 0.10$  \\
\textit{F2 H} & 25.13$\,\pm\, 0.02$ & 25.12$\,\pm\, 0.03$ & 25.02$\,\pm\, 0.03$ \\
\textit{F2 $K_s$} & 24.71$\,\pm\, 0.03$ & 24.36$\,\pm\, 0.02$ & 24.25$\,\pm\, 0.07$  
\enddata
\end{deluxetable*}

\onecolumngrid
\section{Additional Tables and Figures}
The following section contains additional tables to supplement the main text. Table \ref{tab:excess_all} presents all of the observed white dwarfs with \textit{Spitzer}-confirmed infrared excess and Table \ref{tab:no-excess_all} presents all of the white dwarfs observed by \textit{Spitzer} without confirmed excess. For both of the tables, we refer to the Sample of origin as either ``Sample E'' or a subtraction between two other Samples, such that a white dwarf from ``Sample A-B'' originates from ``Sample A'' subtracted by ''Sample B''. The photometry flags, ``g'' and ``s'', show when the Gemini photometry is based on a low number of reference stars and when the Spitzer PRF residual is not clean. Additionally, the $\chi_{\rm{J}}$ flag indicates that the excess is likely due to a companion rather than a dust disk. Both tables are published in their entirety in machine-readable format\footnote{Temporarily available at \url{https://www.mso.anu.edu.au/~samlai/Table_A2_MR.csv} and \url{https://www.mso.anu.edu.au/~samlai/Table_A3_MR.csv}}. The machine-readable versions include the DA white dwarf effective temperature and surface gravity fits from \cite{Gentile_2019WDs}.

\clearpage
\startlongtable
\begin{longtable*}{@{\extracolsep{\fill}}lllccccccccc}
\caption {\label{tab:excess_all} Excess metrics of all observed white dwarfs with \textit{Spitzer}-confirmed excess. \textit{Spitzer} photometry is denoted by ``\textit{S\_Ch1}'' and ``\textit{S\_Ch2}'' for the two warm channels. Photometry flag, ``g'', shows when the Gemini photometry is based on a low number of reference stars and subscripts, ``U'' and ``V'', indicate where \textit{UKIRT} or \textit{VISTA} photometry have been used in place of or in absence of Gemini photometry for the near-infrared $\chi_{\rm{i}}$ measurement. The $\chi_{\rm{J}}$ flag indicates that the excess is likely due to a companion rather than a dust disk.} \\ \hline \hline
 Name  & \textit{Gaia} RA & \textit{Gaia} Dec &\multicolumn{5}{c}{$\chi_{\rm{i}}$} & $\Sigma_{\rm{ch1-ch2}}$ & Flags & Excess & Sample  \\
\cline{4-8} 
      & (deg) & (deg) &  \textit{J}  & \textit{H}  & \textit{K} & \textit{S\_Ch1} & \textit{S\_Ch2}                                        \\    \hline
GaiaJ0006+2858 & 1.644751 & 28.979653 & 0.47 & 1.99 & 7.21 & 25.84 & 33.56 & 5.32 &  & Color+Flux & E \\ 
GaiaJ0007+1951 & 1.948442 & 19.856755 & 8.04 & 13.96 & 11.61 & 24.64 & 24.99 & 0.16 & $\chi_{\rm{J}}$ & FluxOnly & E \\ 
GaiaJ0050-0326 & 12.690832 & -3.448819 & 1.51$_{\rm{V}}$ & 2.39$_{\rm{V}}$ & 1.71$_{\rm{V}}$ & 9.00 & 11.18 & 1.34 &  & FluxOnly & A-B \\ 
GaiaJ0052+4505 & 13.018277 & 45.092720 & 1.95 & 2.70 & 9.97 & 10.50 & 11.32 & 0.53 &  & FluxOnly & E \\ 
GaiaJ0119-7655 & 19.778307 & -76.917482 & 2.60$_{\rm{V}}$ &  & 1.14$_{\rm{V}}$ & 5.08 & 8.97 & 2.61 &  & FluxOnly & E \\ 
GaiaJ0147+2329 & 26.978383 & 23.661691 & 0.33$_{\rm{U}}$ & 0.08 & 5.05 & 18.80 & 24.09 & 3.67 & g & Color+Flux & A-B \\ 
GaiaJ0205-7941 & 31.358532 & -79.684393 & 0.60$_{\rm{V}}$ &  & 1.96$_{\rm{V}}$ & 8.35 & 13.09 & 3.21 &  & Color+Flux & E \\ 
GaiaJ0234-0406 & 38.564633 & -4.102482 & -0.40$_{\rm{V}}$ & -0.01$_{\rm{V}}$ & 1.41$_{\rm{V}}$ & 9.36 & 13.75 & 3.01 &  & Color+Flux & E \\ 
GaiaJ0257+5103 & 44.341677 & 51.062136 & 1.00 & 1.72 & 1.91 & 4.70 & 5.04 & 0.22 &  & FluxOnly & C-D \\ 
GaiaJ0347+1624 & 56.902909 & 16.402432 & 1.87 & 3.62 & 5.94 & 10.50 & 16.98 & 4.53 &  & Color+Flux & E \\ 
GaiaJ0412-4510 & 63.212121 & -45.169625 & 2.36$_{\rm{V}}$ &  & 8.12$_{\rm{V}}$ & 8.74 & 11.03 & 1.58 &  & FluxOnly & E \\ 
GaiaJ0433+2827 & 68.477686 & 28.457861 & 9.01 & 17.01 & 18.48 & 24.05 & 24.86 & 0.48 & $\chi_{\rm{J}}$ & FluxOnly & A-B \\ 
GaiaJ0455+5913 & 73.888221 & 59.222701 & 1.08 & 0.50 & 4.73 & 9.02 & 11.30 & 1.44 &  & FluxOnly & E \\ 
GaiaJ0507+4541 & 76.848342 & 45.695697 & 3.68 & -4.39 & 2.89 & 5.89 & 10.59 & 3.09 & $\chi_{\rm{J}}$g & Color+Flux & E \\ 
GaiaJ0510+2315 & 77.508735 & 23.261340 & -0.22$_{\rm{U}}$ & 1.87 & 10.53$_{\rm{U}}$ & 13.69 & 19.71 & 4.18 & g & Color+Flux & E \\ 
GaiaJ0518+6753 & 79.605945 & 67.897284 & 0.24 & 6.98 & 7.05 & 13.33 & 18.18 & 3.32 & g & Color+Flux & E \\ 
GaiaJ0603+4518 & 90.786308 & 45.307728 & 0.87 & 1.25 & 3.79 & 7.83 & 8.78 & 0.64 &  & FluxOnly & E \\ 
GaiaJ0644-0352 & 101.021923 & -3.868553 & 0.16 & 1.37 & 4.77 & 20.47 & 27.58 & 4.91 &  & Color+Flux & E \\ 
GaiaJ0649-7624 & 102.395218 & -76.416141 & 0.72$_{\rm{V}}$ &  & 4.76$_{\rm{V}}$ & 17.60 & 23.87 & 4.24 &  & Color+Flux & E \\ 
GaiaJ0701+2321 & 105.257690 & 23.365196 & -0.03 & 0.49 &  & 11.32 & 16.47 & 3.48 &  & Color+Flux & E \\ 
GaiaJ0723+6301 & 110.823030 & 63.024058 & -1.07 & -1.15 & 0.39 & 9.08 & 13.80 & 3.14 & g & Color+Flux & E \\ 
GaiaJ0731+2417 & 112.793067 & 24.284180 & -0.42 & -0.55 & 7.18 & 18.62 & 24.85 & 4.29 &  & Color+Flux & E \\ 
GaiaJ0747-0301 & 116.790374 & -3.029560 & 0.68$_{\rm{V}}$ &  & 1.93$_{\rm{V}}$ & 7.70 & 12.74 & 3.36 &  & Color+Flux & E \\ 
GaiaJ0751+1059 & 117.939950 & 10.992025 & 4.19 & 7.40 & 9.21 & 9.13 & 10.23 & 0.87 & $\chi_{\rm{J}}$ & FluxOnly & E \\ 
GaiaJ0802+5631 & 120.615318 & 56.532014 & -0.49 & -1.58 & 2.53 & 11.54 & 16.69 & 3.58 &  & Color+Flux & E \\ 
GaiaJ0832+8149 & 128.140599 & 81.827178 & 1.14 & -0.24 & 3.00 & 16.51 & 25.28 & 6.03 &  & Color+Flux & E \\ 
GaiaJ0841-6511 & 130.439571 & -65.195254 & 4.35$_{\rm{V}}$ &  & 8.62$_{\rm{V}}$ & 12.29 & 12.62 & 0.16 & $\chi_{\rm{J}}$ & FluxOnly & E \\ 
GaiaJ0844+3329 & 131.068177 & 33.487489 & 0.29$_{\rm{U}}$ & -0.05 & 1.68 & 4.57 & 5.04 & 0.27 & g & FluxOnly & E \\ 
GaiaJ0854-7646 & 133.739982 & -76.772772 & -0.88$_{\rm{V}}$ &  & -0.89$_{\rm{V}}$ & 4.94 & 9.79 & 3.17 &  & Color+Flux & E \\ 
GaiaJ0942-1950 & 145.500089 & -19.839612 & 1.04$_{\rm{V}}$ &  & 0.94$_{\rm{V}}$ & 8.90 & 8.93 & -0.12 &  & FluxOnly & D-E \\ 
GaiaJ1030-1435 & 157.582665 & -14.590177 & 1.24 & 4.12 & 3.88 & 11.11 & 11.45 & 0.14 & g & FluxOnly & E \\ 
GaiaJ1039-0325 & 159.922936 & -3.426255 & -1.25$_{\rm{V}}$ & -1.50$_{\rm{V}}$ & -1.70$_{\rm{V}}$ & 5.83 & 6.59 & 0.47 & g & FluxOnly & D-E \\ 
GaiaJ1100-1350 & 165.221455 & -13.839100 & 3.07$_{\rm{V}}$ & 4.32$_{\rm{V}}$ & 6.12$_{\rm{V}}$ & 14.04 & 16.05 & 1.34 & $\chi_{\rm{J}}$ & FluxOnly & E \\ 
GaiaJ1102-1653 & 165.510180 & -16.890932 & 0.96$_{\rm{V}}$ &  & 0.99$_{\rm{V}}$ & 9.75 & 9.56 & -0.17 &  & FluxOnly & E \\ 
GaiaJ1131-1438 & 172.772997 & -14.635475 & 5.88$_{\rm{V}}$ & 13.45$_{\rm{V}}$ & 18.38$_{\rm{V}}$ & 21.96 & 24.34 & 1.66 & $\chi_{\rm{J}}$ & FluxOnly & E \\ 
GaiaJ1146-3636 & 176.618373 & -36.605444 & 0.73$_{\rm{V}}$ &  & 1.61$_{\rm{V}}$ & 7.04 & 15.23 & 5.25 &  & Color+Flux & E \\ 
GaiaJ1155+2649 & 178.775677 & 26.823271 & -0.35 & -0.31 & -1.16 & 4.68 & 6.66 & 1.28 &  & FluxOnly & A-B \\ 
GaiaJ1319+6433 & 199.961921 & 64.552634 &  &  &  & 6.85 & 6.89 & 0.00 &  & FluxOnly & E \\ 
GaiaJ1322-1210 & 200.748077 & -12.178491 & 0.49$_{\rm{V}}$ & 1.47$_{\rm{V}}$ & -0.14$_{\rm{V}}$ & 3.61 & 4.91 & 0.88 &  & FluxOnly & D-E \\ 
GaiaJ1343-0453 & 205.859499 & -4.896320 & 0.12$_{\rm{V}}$ & -0.64$_{\rm{V}}$ & -0.78$_{\rm{V}}$ & -0.93 & 5.45 & 4.34 &  & ColorOnly & C-D \\ 
GaiaJ1456+1704 & 224.171323 & 17.070933 & -0.17 & -0.83 & -1.70 & 2.82 & 11.54 & 5.46 &  & ColorOnly & E \\ 
GaiaJ1539-3910 & 234.821649 & -39.180964 & 0.98$_{\rm{V}}$ &  & 5.06$_{\rm{V}}$ & 9.83 & 11.07 & 0.80 &  & FluxOnly & C-D \\ 
GaiaJ1613+5521 & 243.319096 & 55.357181 & 0.30$_{\rm{U}}$ & 0.41 & 3.97$_{\rm{U}}$ & 7.86 & 11.30 & 2.38 & g & FluxOnly & E \\ 
GaiaJ1622+5840 & 245.748880 & 58.674695 & -0.54$_{\rm{U}}$ & -1.04 & 2.99 & 12.95 & 19.43 & 4.53 & g & Color+Flux & E \\ 
GaiaJ1728+2053 & 262.190381 & 20.894718 & -0.03 & -0.61 & -0.46 & 5.90 & 9.52 & 2.45 &  & FluxOnly & D-E \\ 
GaiaJ1731-1002 & 262.771413 & -10.036248 & 18.45$_{\rm{V}}$ &  & 39.85$_{\rm{V}}$ & 23.73 & 22.75 & -0.95 & $\chi_{\rm{J}}$ & FluxOnly & A-B \\ 
GaiaJ1814-7355 & 273.573350 & -73.917388 & 0.93 & 1.46 & 5.28 & 24.57 & 32.05 & 5.19 &  & Color+Flux & A-B \\ 
GaiaJ1903+6035 & 285.833014 & 60.598328 & -0.84 & -1.58 & 1.65 & 14.28 & 21.24 & 4.86 &  & Color+Flux & E \\ 
GaiaJ1939+0932 & 294.979609 & 9.538653 & -0.35$_{\rm{U}}$ &  &  & 18.58 & 20.20 & 0.93 &  & FluxOnly & A-B \\ 
GaiaJ1949+7007 & 297.446756 & 70.121605 &  &  &  & 13.01 & 14.89 & 1.49 &  & FluxOnly & E \\ 
GaiaJ2015+5531 & 303.861206 & 55.520607 & 0.31$_{\rm{U}}$ &  &  & 5.77 & 12.14 & 4.24 &  & Color+Flux & A-B \\ 
GaiaJ2026+5925 & 306.588176 & 59.423365 & 11.00$_{\rm{U}}$ &  &  & 16.52 & 14.78 & -1.30 & $\chi_{\rm{J}}$ & FluxOnly & C-D \\ 
GaiaJ2044-7842 & 311.143154 & -78.700513 & 0.96 & 2.32 & 1.55 & 3.42 & 5.53 & 1.38 &  & FluxOnly & E \\ 
GaiaJ2048+1333 & 312.191021 & 13.565169 & 7.93 & 17.97 & 11.71 & 21.81 & 22.34 & 0.35 & $\chi_{\rm{J}}$ & FluxOnly & E \\ 
GaiaJ2100+2122 & 315.144710 & 21.382635 & 0.45 & 2.47 & 7.72 & 24.72 & 31.45 & 4.66 &  & Color+Flux & E \\ 
GaiaJ2155+7610 & 328.763388 & 76.169309 & 0.86 & -0.35 & 2.30 & 7.34 & 12.54 & 3.50 &  & Color+Flux & E \\ 
GaiaJ2205-4610 & 331.303399 & -46.180975 & 0.93$_{\rm{V}}$ & 1.16$_{\rm{V}}$ & 1.15$_{\rm{V}}$ & 4.88 & 5.78 & 0.57 &  & FluxOnly & E \\ 
GaiaJ2223-2510 & 335.993387 & -25.178781 & -0.89$_{\rm{V}}$ &  & 1.08$_{\rm{V}}$ & 12.92 & 20.01 & 4.89 &  & Color+Flux & B-C \\ 
GaiaJ2248-0642 & 342.166938 & -6.712539 & 0.66$_{\rm{V}}$ & 0.77$_{\rm{V}}$ & 0.05$_{\rm{V}}$ & 6.42 & 4.66 & 1.53 &  & FluxOnly & E \\ 
GaiaJ2253+0833 & 343.333391 & 8.561881 & 0.63$_{\rm{U}}$ & 1.00$_{\rm{U}}$ & 0.24$_{\rm{U}}$ & 8.17 & 5.70 & -1.73 &  & FluxOnly & E \\ 
GaiaJ2306+2702 & 346.727225 & 27.035886 & -1.51$_{\rm{U}}$ & 0.48 & 0.32 & 11.13 & 17.58 & 4.35 & g & Color+Flux & E \\ 
GaiaJ2330+2934 & 352.654068 & 29.577891 & 0.28$_{\rm{U}}$ & -0.41 & 0.35 & 4.08 & 7.68 & 2.39 & g & FluxOnly & A-B \\ 
 \hline
\multicolumn{6}{l}{\footnotesize
$^{\rm{g}}$Gemini photometry flag. 
$^{\chi_{\rm{J}}}$ Stellar companion flag.}
\end{longtable*}
\clearpage

\clearpage
\startlongtable
\begin{longtable*}{@{\extracolsep{\fill}}lllcccccccc}
\caption {\label{tab:no-excess_all} Excess metrics of all white dwarfs observed with \textit{Spitzer} without infrared excess. Spitzer photometry is denoted by ``\textit{S\_Ch1}'' and ``\textit{S\_Ch2}'' for the two warm channels. Photometry flags, ``g'' and ``s'', show when the Gemini photometry is based on a low number of reference stars and when the Spitzer photometry is unreliable. Subscripts, ``U'' and ``V'', indicate where \textit{UKIRT} or \textit{VISTA} photometry have been used in place of or in absence of Gemini photometry for the near-infrared $\chi_{\rm{i}}$ measurement.} \\ \hline \hline
 Name  & \textit{Gaia} RA & \textit{Gaia} Dec &\multicolumn{5}{c}{$\chi_{\rm{i}}$} & $\Sigma_{\rm{ch1-ch2}}$ & Flags  & Sample  \\
\cline{4-8} 
      & (deg) & (deg) &  \textit{J}  & \textit{H}  & \textit{K} & \textit{S\_Ch1} & \textit{S\_Ch2}                                        \\    \hline
GaiaJ0055+1135 & 13.892676 & 11.583566 & -0.47$_{\rm{U}}$ & -0.77$_{\rm{U}}$ & 0.79$_{\rm{U}}$ & -0.40 & -2.56 & -1.56 &  & D-E \\ 
GaiaJ0104+3816 & 16.080300 & 38.281834 & -0.20$_{\rm{U}}$ &  &  & -0.60 & 0.45 & 0.74 &  & E \\ 
GaiaJ0107+2518 & 16.859511 & 25.309778 & 0.48 & 0.25 & -0.45 & 0.35 & 2.24 & 1.30 &  & D-E \\ 
GaiaJ0111+3136 & 17.987846 & 31.608898 & 1.13$_{\rm{U}}$ & 0.34 & -1.89 & 1.16 & 1.31 & 0.08 & g & C-D \\ 
GaiaJ0133+0816 & 23.347031 & 8.267714 & -0.08$_{\rm{U}}$ & -0.66$_{\rm{U}}$ & -1.73$_{\rm{U}}$ & 0.10 & 0.91 & 0.57 &  & A-B \\ 
GaiaJ0151-2503 & 27.998720 & -25.054374 & -0.31 & -1.85 & 2.31 & -1.15 & -1.78 & -0.42 & g & A-B \\ 
GaiaJ0206-2316 & 31.688827 & -23.271080 & 0.44$_{\rm{V}}$ & -0.52 & 0.22$_{\rm{V}}$ & 0.16 & 0.46 & 0.21 & g & D-E \\ 
GaiaJ0256+2334 & 44.074243 & 23.573071 & 0.09$_{\rm{U}}$ & -1.19 & -0.56 & 0.58 & 1.17 & 0.41 & g & D-E \\ 
GaiaJ0258-1048 & 44.590649 & -10.807481 & 0.17$_{\rm{V}}$ &  & -0.62$_{\rm{V}}$ & -0.41 & 0.22 & 0.45 &  & D-E \\ 
GaiaJ0323-5030 & 50.980972 & -50.506115 & 0.51$_{\rm{V}}$ & 1.10$_{\rm{V}}$ & -1.04$_{\rm{V}}$ & 1.55 & 1.69 & 0.05 &  & D-E \\ 
GaiaJ0328+2528 & 52.071458 & 25.481511 & 0.26 & 0.61 & 0.45 & 5.22 & 4.25 & -0.69 &  & D-E \\ 
GaiaJ0329-4738 & 52.363076 & -47.644224 & 1.72$_{\rm{V}}$ & 1.09$_{\rm{V}}$ & 0.50$_{\rm{V}}$ & 1.21 & 1.29 & 0.05 &  & D-E \\ 
GaiaJ0329-5346 & 52.430831 & -53.767221 & 0.73$_{\rm{V}}$ &  & -0.54$_{\rm{V}}$ & -0.40 & -0.32 & 0.06 &  & A-B \\ 
GaiaJ0346+1247 & 56.743468 & 12.791663 & 2.08$_{\rm{U}}$ & 0.73 & -0.58 & 2.62 & 2.80 & 0.10 & g & A-B \\ 
GaiaJ0348+5150 & 57.146356 & 51.838514 & 0.10 & -0.20 & 0.42 & 0.74 & 0.46 & -0.20 &  & D-E \\ 
GaiaJ0359-2154 & 59.852820 & -21.905147 & 0.98 & -0.44 & -1.25 & -0.83 & -0.33 & 0.37 &  & D-E \\ 
GaiaJ0404+1502 & 61.146002 & 15.040344 & 1.51$_{\rm{U}}$ & 1.23 & 1.33$_{\rm{U}}$ & 1.47 & 2.98 & 1.05 & g & E \\ 
GaiaJ0413-1235 & 63.438697 & -12.594478 & -0.31$_{\rm{V}}$ &  & -0.48$_{\rm{V}}$ & 1.24 & 1.47 & 0.16 &  & E \\ 
GaiaJ0421+1529 & 65.453585 & 15.487452 & 1.48 & 1.56 & 1.93 & 3.38 & 4.49 & 0.74 &  & D-E \\ 
GaiaJ0424+0348 & 66.067091 & 3.814391 & 0.99 & 2.11 & 1.88 & 1.72 & 2.04 & 0.18 &  & A-B \\ 
GaiaJ0428+3644 & 67.078300 & 36.739478 & -0.12 & -0.75 & 0.06 & 0.66 & 4.97 & 2.91 &  & D-E \\ 
GaiaJ0518-0757 & 79.633570 & -7.955206 & 1.09$_{\rm{V}}$ &  & -1.12$_{\rm{V}}$ & 0.45 & 1.01 & 0.39 &  & D-E \\ 
GaiaJ0528-6442 & 82.175796 & -64.708519 & 0.64 & 1.61 & 3.04 & 4.16 & 4.13 & -0.05 &  & A-B \\ 
GaiaJ0531-4557 & 82.752770 & -45.966459 & 1.03$_{\rm{V}}$ &  & -0.49$_{\rm{V}}$ & 0.43 & 1.15 & 0.49 &  & D-E \\ 
GaiaJ0536-3254 & 84.167638 & -32.915170 & 0.03 & -0.14 & 0.43 & -0.86 & -0.87 & 0.03 &  & D-E \\ 
GaiaJ0609+3913 & 92.250011 & 39.222588 & 0.26 & -0.14 & 0.39 & 2.10 & 1.93 & -0.13 &  & E \\ 
GaiaJ0620+3443 & 95.162010 & 34.718149 & -0.22 & -1.33 & -0.61 & -0.17 & 1.18 & 0.95 &  & D-E \\ 
GaiaJ0639+6147 & 99.993800 & 61.789765 & -2.36 & 0.57 & -4.04 & 1.41 & 0.43 & -0.70 & g & E \\ 
GaiaJ0707+2651 & 106.979289 & 26.850791 & 0.53 & 0.88 & 0.58 & 0.11 & 1.80 & 1.18 &  & C-D \\ 
GaiaJ0711+0928 & 107.932042 & 9.480980 & -0.47$_{\rm{U}}$ &  &  & -0.33 & 2.31 & 1.84 &  & E \\ 
GaiaJ0723+1617 & 110.750828 & 16.284668 & 0.34$_{\rm{U}}$ & 0.00 & -0.20 & -0.03 & 0.26 & 0.21 & g & D-E \\ 
GaiaJ0831+7155 & 127.833370 & 71.927980 & 0.12 & -1.74 & -0.31 & 2.93 & 7.72 & 2.99 &  & E \\ 
GaiaJ0834+5336 & 128.588430 & 53.604311 & 0.44 & -1.69 & -1.26 & 0.63 & 0.38 & -0.19 &  & A-B \\ 
GaiaJ0834-3450 & 128.732977 & -34.847287 & 0.82 & 0.69 & 0.34 & 0.55 & 0.82 & 0.18 & g & A-B \\ 
GaiaJ0842+3748 & 130.577991 & 37.816440 & -0.45$_{\rm{U}}$ & -1.76 & 0.50 & 0.80 & 1.79 & 0.68 & g & C-D \\ 
GaiaJ0844-4408 & 131.232031 & -44.142421 & 0.87 & -0.35 & 0.59 & 0.56 & 0.98 & 0.29 & g & A-B \\ 
GaiaJ0845+0653 & 131.407110 & 6.896151 & -0.75$_{\rm{U}}$ & -0.94$_{\rm{U}}$ & -2.00$_{\rm{U}}$ & 0.73 & -0.12 & -0.61 &  & A-B \\ 
GaiaJ0845+6009 & 131.462714 & 60.153801 &  &  &  & 0.40 & 0.29 & -0.07 &  & D-E \\ 
GaiaJ0847-1859 & 131.871782 & -18.997004 & 0.04$_{\rm{V}}$ &  & -0.95$_{\rm{V}}$ & -0.69 & -1.29 & -0.41 &  & D-E \\ 
GaiaJ0902+3120 & 135.677408 & 31.345378 & -0.18$_{\rm{U}}$ & 0.12$_{\rm{U}}$ & -0.05$_{\rm{U}}$ & 0.66 & 0.89 & 0.15 & g & D-E \\ 
GaiaJ0904+5935 & 136.246569 & 59.588664 & -0.81$_{\rm{U}}$ & -0.61 & -0.89 & 2.10 & 2.51 & 0.30 & g & A-B \\ 
GaiaJ0906+2836 & 136.611269 & 28.601867 & -0.04$_{\rm{U}}$ & 0.74$_{\rm{U}}$ & 0.87$_{\rm{U}}$ & 0.66 & 1.38 & 0.46 &  & A-B \\ 
GaiaJ0936-3721 & 144.247925 & -37.356694 & 0.70$_{\rm{V}}$ &  & -0.51$_{\rm{V}}$ & 0.08 & 0.39 & 0.22 &  & D-E \\ 
GaiaJ0940+1903 & 145.034226 & 19.065272 & -0.34$_{\rm{U}}$ & -0.26 & -0.67 & 0.29 & 3.30 & 2.02 & g & E \\ 
GaiaJ0947+2616 & 146.861039 & 26.267282 & 0.15$_{\rm{U}}$ & 0.02 & -0.55 & -1.15 & -1.30 & -0.13 & g & A-B \\ 
GaiaJ0950+1837 & 147.529107 & 18.625792 & -0.62 & -0.90 & -1.38 & 0.29 & -0.52 & -0.58 &  & A-B \\ 
GaiaJ0959-1135 & 149.755167 & -11.590022 & -0.55$_{\rm{V}}$ & 0.25$_{\rm{V}}$ & -0.72$_{\rm{V}}$ & -0.47 & -0.45 & 0.03 &  & A-B \\ 
GaiaJ1001-0842 & 150.388500 & -8.714054 & 1.83 & 1.15 & 0.01 & 0.67 & 1.21 & 0.36 &  & B-C \\ 
GaiaJ1017-3236 & 154.368509 & -32.602488 & 0.20$_{\rm{V}}$ &  & -0.00$_{\rm{V}}$ & 1.28 & 0.97 & -0.22 &  & A-B \\ 
GaiaJ1040+2848 & 160.218776 & 28.815718 & 0.13$_{\rm{U}}$ & -1.48 & -0.16 & 1.44 & 2.81 & 0.93 & g & A-B \\ 
GaiaJ1104+2356 & 166.157916 & 23.944726 & -0.39$_{\rm{U}}$ & 0.08 & -0.47 & 0.65 & 2.02 & 0.95 & g & A-B \\ 
GaiaJ1122+6711 & 170.566707 & 67.196167 &  &  &  & 0.61 & 1.11 & 0.36 &  & A-B \\ 
GaiaJ1125+4223 & 171.423117 & 42.392897 & 0.34$_{\rm{U}}$ &  &  & 4.81 & 3.28 & -1.09 &  & A-B \\ 
GaiaJ1136-3807 & 174.085921 & -38.127104 & 0.97$_{\rm{V}}$ &  & -0.19$_{\rm{V}}$ & 2.96 & 4.22 & 0.83 &  & C-D \\ 
GaiaJ1218+2648 & 184.690397 & 26.808761 & -0.47$_{\rm{U}}$ & 1.41$_{\rm{U}}$ & -1.18$_{\rm{U}}$ & -0.05 & 0.19 & 0.17 & g & D-E \\ 
GaiaJ1226-6612 & 186.674734 & -66.205917 & -0.36 & -1.08 & -0.73 & 0.26 & 0.38 & 0.09 & g & A-B \\ 
GaiaJ1252+0410 & 193.063270 & 4.178632 & -0.04$_{\rm{U}}$ & 1.27$_{\rm{U}}$ & 0.55$_{\rm{U}}$ & 2.65 & 4.09 & 1.00 &  & D-E \\ 
GaiaJ1257-4646 & 194.317310 & -46.780228 & 1.16$_{\rm{V}}$ &  & 0.37$_{\rm{V}}$ & 0.11 & 0.70 & 0.41 &  & A-B \\ 
GaiaJ1307-1017 & 196.950001 & -10.299630 & 0.27$_{\rm{V}}$ & -0.20$_{\rm{V}}$ & -1.71$_{\rm{V}}$ & 1.56 & 4.97 & 2.25 &  & D-E \\ 
GaiaJ1312-3733 & 198.246288 & -37.565289 & 1.19$_{\rm{V}}$ &  & 0.67$_{\rm{V}}$ & 1.62 & 1.35 & -0.20 &  & C-D \\ 
GaiaJ1350+2434 & 207.634955 & 24.570743 & 0.08$_{\rm{U}}$ & -0.87$_{\rm{U}}$ & -1.84$_{\rm{U}}$ & 2.37 & 2.88 & 0.34 &  & D-E \\ 
GaiaJ1354+0108 & 208.748953 & 1.138578 & -0.01$_{\rm{U}}$ & 0.33$_{\rm{U}}$ & 0.28$_{\rm{U}}$ & 1.14 & 2.31 & 0.81 &  & A-B \\ 
GaiaJ1424+0444 & 216.107439 & 4.743526 & -1.08$_{\rm{U}}$ & 0.19$_{\rm{U}}$ & 0.31$_{\rm{U}}$ & 3.10 & 4.54 & 0.91 &  & A-B \\ 
GaiaJ1429-2751 & 217.363636 & -27.850811 & 1.62$_{\rm{V}}$ &  & 0.90$_{\rm{V}}$ & 0.08 & 0.88 & 0.55 &  & D-E \\ 
GaiaJ1434+1508 & 218.528056 & 15.138231 & -0.10$_{\rm{U}}$ & -0.43 & -3.13 & 0.85 & 1.51 & 0.46 & g & A-B \\ 
GaiaJ1449-3029 & 222.388494 & -30.488730 & -1.42 & -0.53 & 1.72 & 2.71 & 3.25 & 0.36 &  & C-D \\ 
GaiaJ1450+4055 & 222.527477 & 40.925948 & 0.21$_{\rm{U}}$ & 1.55 & 1.66 & 1.35 & 3.20 & 1.29 & g & E \\ 
GaiaJ1510-4143 & 227.720492 & -41.732524 & -0.29$_{\rm{V}}$ &  & 0.17$_{\rm{V}}$ & 0.52 & 1.59 & 0.74 &  & A-B \\ 
GaiaJ1516-3545 & 229.088293 & -35.761958 & 1.68$_{\rm{V}}$ &  & 1.76$_{\rm{V}}$ & 2.59 & 3.02 & 0.28 &  & C-D \\ 
GaiaJ1518-1148 & 229.619733 & -11.811128 & 1.13$_{\rm{V}}$ & 1.63$_{\rm{V}}$ & -0.47$_{\rm{V}}$ & 3.18 & 5.58 & 1.64 &  & E \\ 
GaiaJ1528-0128 & 232.206188 & -1.482796 & 0.08$_{\rm{U}}$ & -0.22$_{\rm{U}}$ & -0.20$_{\rm{U}}$ & 0.31 & -0.27 & -0.41 &  & A-B \\ 
GaiaJ1532+4231 & 233.192486 & 42.527308 & 0.72 & 1.01 & 1.34 & -0.60 & 0.84 & 1.01 & g & A-B \\ 
GaiaJ1539-7225 & 234.816900 & -72.430662 & 1.66 & -0.08 & 0.58 & 0.82 & 1.55 & 0.50 &  & D-E \\ 
GaiaJ1546-0557 & 236.672823 & -5.955513 & 1.40$_{\rm{V}}$ & 2.20$_{\rm{V}}$ & 2.37$_{\rm{V}}$ & 3.12 & 6.80 & 2.55 &  & D-E \\ 
GaiaJ1548+2451 & 237.229549 & 24.853610 & 0.08$_{\rm{U}}$ & -0.82 & 0.85 & 1.73 & 0.79 & -0.68 & g & A-B \\ 
GaiaJ1612+1419 & 243.026837 & 14.318464 & 0.00$_{\rm{U}}$ & 0.05 & -0.60 & 0.82 & 1.00 & 0.12 & g & C-D \\ 
GaiaJ1632-2058 & 248.033976 & -20.969509 & 0.67$_{\rm{V}}$ & 0.69$_{\rm{V}}$ & 0.13$_{\rm{V}}$ & 1.24 & 2.13 & 0.63 &  & A-B \\ 
GaiaJ1634+2812 & 248.546838 & 28.203436 & -0.09$_{\rm{U}}$ & -0.13$_{\rm{U}}$ & -0.22$_{\rm{U}}$ & -0.35 & -0.11 & 0.22 &  & A-B \\ 
GaiaJ1635+1343 & 248.907987 & 13.733185 & 1.36 & 2.13 & 0.91 & 3.07 & 4.16 & 0.67 &  & C-D \\ 
GaiaJ1702+5034 & 255.578581 & 50.582736 & 0.86$_{\rm{U}}$ &  &  & 3.46 & 4.42 & 0.64 &  & C-D \\ 
GaiaJ1706-7623 & 256.604920 & -76.384867 & 0.58 &  & 0.59 & 2.41 & 2.54 & 0.08 &  & A-B \\ 
GaiaJ1729+5010 & 262.443317 & 50.167743 & 0.54$_{\rm{U}}$ & -1.38 & 2.90 & -0.00 & 0.21 & 0.15 & g & D-E \\ 
GaiaJ1745-1317 & 266.477577 & -13.298303 & 2.28$_{\rm{V}}$ &  & 3.24$_{\rm{V}}$ & 1.69 & 1.69 & -0.00 &  & A-B \\ 
GaiaJ1755+3958 & 268.980820 & 39.978672 & 1.32 &  & 1.75 & 0.50 & 1.15 & 0.44 &  & E \\ 
GaiaJ1820+7454 & 275.155315 & 74.900837 &  &  &  & 1.72 & 2.82 & 0.76 &  & D-E \\ 
GaiaJ1832+7116 & 278.032579 & 71.280554 &  &  &  & -1.38 & -0.77 & 0.47 &  & A-B \\ 
GaiaJ1849-0957 & 282.449430 & -9.963228 & 0.17 & 0.36 & 0.06 & 0.97 & 1.65 & 0.48 & g & A-B \\ 
GaiaJ1858-8432 & 284.586533 & -84.543430 & 1.04$_{\rm{V}}$ &  & 0.38$_{\rm{V}}$ & 1.58 & 1.96 & 0.22 &  & A-B \\ 
GaiaJ1932-5135 & 293.096890 & -51.586555 & -0.07$_{\rm{V}}$ &  &  & 0.43 & 1.74 & 0.88 &  & A-B \\ 
GaiaJ1949-3147 & 297.335513 & -31.786286 & 1.37$_{\rm{V}}$ &  & 0.83$_{\rm{V}}$ & 1.06 & 0.52 & -0.38 &  & D-E \\ 
GaiaJ2012-5957 & 303.135383 & -59.953679 & -5.66$_{\rm{V}}$ &  & -13.51$_{\rm{V}}$ & -17.26 & -18.81 & -1.03 &  & A-B \\ 
GaiaJ2051-7538 & 312.792373 & -75.640037 & 1.39$_{\rm{V}}$ & -1.86 & 1.23$_{\rm{V}}$ & 0.98 & 1.81 & 0.57 & g & D-E \\ 
GaiaJ2103-1729 & 315.966770 & -17.490886 & 0.69$_{\rm{V}}$ &  & -0.80$_{\rm{V}}$ & 1.58 & 1.50 & -0.06 &  & C-D \\ 
GaiaJ2105-4255 & 316.441572 & -42.917937 & 0.78$_{\rm{V}}$ &  & -0.17$_{\rm{V}}$ & 2.15 & 3.65 & 0.99 &  & A-B \\ 
GaiaJ2109+6507 & 317.468728 & 65.122863 & -0.70 & -1.07 & 1.39 & -0.08 & -0.31 & -0.16 &  & D-E \\ 
GaiaJ2110+1746 & 317.748107 & 17.775631 & 0.71$_{\rm{U}}$ & 0.38 & 0.02 & 2.55 & 3.31 & 0.51 & g & A-B \\ 
GaiaJ2124+8556 & 321.082404 & 85.946076 &  &  &  & -0.48 & 0.27 & 0.53 &  & D-E \\ 
GaiaJ2152-7207 & 328.225832 & -72.118972 & 0.57 & -0.82 & -0.20 & -0.27 & 0.50 & 0.55 &  & D-E \\ 
GaiaJ2202+2919 & 330.523145 & 29.318463 & -0.73 & -0.70 & -0.40 & 0.61 & 1.41 & 0.56 &  & C-D \\ 
GaiaJ2213-5020 & 333.340992 & -50.334322 & 0.60$_{\rm{V}}$ & 0.55$_{\rm{V}}$ & -0.23$_{\rm{V}}$ & 1.80 & 2.24 & 0.27 &  & C-D \\ 
GaiaJ2225-1125 & 336.254028 & -11.427853 & 0.67$_{\rm{V}}$ &  & -1.07$_{\rm{V}}$ & 0.57 & -1.25 & -1.28 &  & A-B \\ 
GaiaJ2233+8408 & 338.321327 & 84.137396 & 0.40 & 0.18 & 0.45 & 1.20 & 0.46 & -0.52 &  & D-E \\ 
GaiaJ2233-3832 & 338.475669 & -38.544736 & 0.20$_{\rm{V}}$ &  & -0.07$_{\rm{V}}$ & 1.41 & 1.79 & 0.26 &  & A-B \\ 
GaiaJ2243+2201 & 340.808435 & 22.024626 & 0.65$_{\rm{U}}$ & -0.27 & -0.44 & -0.19 & -0.00 & 0.14 & g & A-B \\ 
GaiaJ2250+3231 & 342.601220 & 32.528655 & -0.13 & -2.18 & 0.04 & -0.50 & 0.57 & 0.76 &  & E \\ 
GaiaJ2255-4405 & 343.917598 & -44.090303 & 1.16$_{\rm{V}}$ &  & 0.27$_{\rm{V}}$ & 2.66 & 5.17 & 1.69 &  & A-B \\ 
GaiaJ2305+5125 & 346.382172 & 51.422359 & -0.50 & -1.44 & 1.56 & 0.14 & 1.82 & 1.18 &  & A-B \\ 
GaiaJ2305+7543 & 346.485706 & 75.731470 & -0.47 & -1.06 & 0.46 & 1.61 & 2.12 & 0.31 &  & A-B \\ 
GaiaJ2332-3301 & 353.045433 & -33.018883 & 0.24$_{\rm{V}}$ & 1.34$_{\rm{V}}$ & 0.32$_{\rm{V}}$ & 2.75 & 7.20 & 2.97 &  & D-E \\ 
GaiaJ2333+0613 & 353.272373 & 6.219436 & 0.46$_{\rm{U}}$ & -0.22$_{\rm{U}}$ & 1.98$_{\rm{U}}$ & -0.83 & 0.53 & 0.97 &  & D-E \\
GaiaJ2349-0616 & 357.481725 & -6.267962 & 0.06$_{\rm{V}}$ & -0.34$_{\rm{V}}$ & -0.46$_{\rm{V}}$ & 1.34 & 1.16 & -0.14 &  & A-B \\ 
GaiaJ2352-0253 & 358.135752 & -2.885370 & -0.01$_{\rm{V}}$ & 0.27$_{\rm{V}}$ & -0.33$_{\rm{V}}$ & 0.72 & 4.76 & 2.75 &  & D-E \\ 
\hline
GaiaJ0106+5604 & 16.585935 & 56.082154 & -1.45$_{\rm{U}}$ &  &  & 5.52 & 4.43 & -0.78 & s & C-D \\ 
GaiaJ1000+6811 & 150.168282 & 68.198294 & 2.57 & 3.17 & 8.67 & 13.11 & 14.92 & 1.11 & s & E \\ 
GaiaJ1046+3745 & 161.748711 & 37.765655 & 1.20 & 1.32 & 5.85 & 12.14 & 17.93 & 3.79 & s & E \\ 
GaiaJ1054+2203 & 163.542967 & 22.053685 & 3.34 & 2.44 & 4.22 & 13.44 & 13.76 & 0.16 & $\chi_{\rm{J}}$s & A-B \\ 
GaiaJ1101+1741 & 165.403567 & 17.698863 & 2.31 & 3.24 & 4.33 & 8.92 & 8.17 & -0.59 & s & A-B \\ 
GaiaJ1102-4921 & 165.738528 & -49.352985 & -0.67 & -1.57 & -0.87 & 3.54 & 5.86 & 1.54 & s & A-B \\ 
GaiaJ1520-0354 & 230.191394 & -3.914617 & -0.12 & -0.70 & -0.58 & 0.92 & 3.34 & 1.66 & s & D-E \\ 
GaiaJ1941-1222 & 295.438744 & -12.381575 & 1.84$_{\rm{V}}$ &  & 1.04$_{\rm{V}}$ & 1.74 &  &  & s & E \\ 
GaiaJ2222-1542 & 335.645848 & -15.703907 & -0.16$_{\rm{V}}$ &  & 0.70$_{\rm{V}}$ & 4.60 & 5.75 & 0.75 & s & D-E \\ 
\hline
\multicolumn{6}{l}{\footnotesize 
$^{\rm{g}}$Gemini photometry flag. 
$^{\rm{s}}$\textit{Spitzer} photometry flag. 
$^{\chi_{\rm{J}}}$ Stellar companion flag.}
\end{longtable*}
\clearpage

\end{document}